\documentclass{aastex631}

\usepackage{graphicx}
\usepackage{amsmath}
\usepackage{savesym}
\savesymbol{tablenum}
\usepackage{siunitx}
\restoresymbol{SIX}{tablenum}
\usepackage{txfonts}

\def\asec{\ifmmode ^{\prime\prime}\else$^{\prime\prime}$\fi}

\def\Msun{\hbox{~{\rm M}_\odot}}

\def\degs{\ifmmode ^{\circ}\else$^{\circ}$\fi}
\def\amin{\ifmmode ^{\prime}\else$^{\prime}$\fi}
\def\asec{\ifmmode ^{\prime\prime}\else$^{\prime\prime}$\fi}
\def\farcs{\hbox{$.\!\!^{\prime\prime}$}}  
\def\h{\hbox{$^{\rm h}$}}
\def\m{\hbox{$^{\rm m}$}}
\def\s{\hbox{$^{\rm s}$}}

\def\degs{\ifmmode ^{\circ}\else$^{\circ}$\fi}
\def\amin{\ifmmode ^{\prime}\else$^{\prime}$\fi}
\def\farcm{\hbox{$.\mkern-4mu^\prime$}}
\def\EE#1{\times 10^{#1}}

\def\cm{\mbox{\,cm}}

\def\cm3{\mbox{\,cm$^{-3}$}}
\def\kms{\mbox{\,km~s$^{-1}$}}

\def\kms{\mbox{\,km s$^{-1}$}}

\def\s{\mbox{\,s}}

\def\lsim{\!\!\!\phantom{\le}\smash{\buildrel{}\over
 {\lower2.5dd\hbox{$\buildrel{\lower2dd\hbox{$\displaystyle<$}}\over
                                 \sim$}}}\,\,}
\def\gsim{\!\!\!\phantom{\ge}\smash{\buildrel{}\over
{\lower2.5dd\hbox{$\buildrel{\lower2dd\hbox{$\displaystyle>$}}\over
                               \sim$}}}\,\,}
\def\Msun{~{\rm M}_\odot}

\begin{document}

\title{Radio studies of supernovae 1979C, 1986J and 2006X with LOFAR}

\author[0000-0001-9896-6994]{Peter Lundqvist}
\affiliation{The Oskar Klein Centre, Department of Astronomy, Stockholm University, AlbaNova, SE-10691 Stockholm, Sweden \\}
\correspondingauthor{Peter Lundqvist}
\email{peteru@astro.su.se}

\author[0000-0002-3664-8082]{Deepika Venkattu}
\affiliation{The Oskar Klein Centre, Department of Astronomy, Stockholm University, AlbaNova, SE-10691 Stockholm, Sweden \\}

\author[0000-0001-5654-0266]{Miguel Pérez Torres}
\affiliation{Instituto de Astrofísica de Andalucía, Glorieta de la Astronomía, s/n, E-18008 Granada, Spain \\
}
\affiliation{Facultad de Ciencias, Universidad de Zaragoza, Pedro Cerbuna 12, E-50009 Zaragoza, Spain \\}

\author[0000-0002-8079-7608]{Javier Moldón}
\affiliation{Instituto de Astrofísica de Andalucía, Glorieta de la Astronomía, s/n, E-18008 Granada, Spain \\
}

\author[0000-0001-5221-2636]{Vijay Mahatma}
\affiliation{Cavendish Laboratory- Astrophysics Group, University of Cambridge, 19 JJ Thomson Avenue, Cambridge, CB3 0HE, United Kingdom\\}
\affiliation{Kavli Institute for Cosmology, University of Cambridge, Madingley Road, Cambridge, CB3 0HA, United Kingdom\\}

\author[0000-0002-0844-6563]{Poonam Chandra}
\affiliation{National Radio Astronomy Observatory, 520 Edgemont Road, Charlottesville, VA 22903, USA \\}

\begin{abstract}
We present LOw Frequency ARray (LOFAR) studies of supernovae SN 1979C, SN 1986J, and SN 2006X, focusing on new observations from the LOFAR Two-metre Sky Survey (LoTSS) and the International LOFAR Telescope (ILT). For Type Ia SN 2006X, we derive a 3$\sigma$ upper limit of 0.7 mJy at 0.146 GHz, and using radio emission models based on the CS15DD2 explosion model, we constrain the circumstellar density to $n_{\rm H} \lsim 10 \cm3$ for the microphysical parameters $\epsilon_{\rm rel} = \epsilon_{\rm B} = 0.01$. 
SN 1979C is clearly detected in the LoTSS image with a flux density of $4.6 \pm 0.36$ mJy nearly 40 years postexplosion. Modeling its radio evolution suggests a steep flux decay ($F_{\nu} \propto t^{-2.1}$) between 22 and 42 years, a break in the spectrum near 1.5 GHz possibly due to synchrotron cooling, a progenitor mass of $\sim 13 \Msun$, and a progressive steepening with velocity for the density slope of the supernova ejecta. Our findings for SN 1979C contradict scenarios involving central compact object emission, and we obtain X-ray temperatures close to those derived from recent observations.
For SN 1986J, we present the first ILT image showing a flux density of $6.77\pm0.2$ mJy at 0.146 GHz. The spectral index of the shell emission is found to be $0.66\pm0.03$, consistent with previous estimates, although variations at low frequencies warrant further investigation. Our results highlight the power of LOFAR for studying long-term radio evolution in supernovae.
\end{abstract}


\section{Introduction} \label{sec:intro}
With the upcoming Square Kilometre Array (SKA) era, there is increased focus on the low-frequency radio sky in the MHz range. In the radio, we typically study synchrotron emission resulting from the interaction of the supernova ejecta with the surrounding circumtellar matter. The study of supernovae remains unexplored in the low-frequency ranges which LOFAR operates in (i.e., between 10 and 240 MHz), opening up a completely uncharted territory for supernovae science. In this work, we choose two nearby galaxies to focus on three supernovae in them. 

M100 (or NGC 4321) is the brightest, and one of the largest spiral galaxies (type SAB(s)bc) in the Virgo cluster. It is a starburst galaxy with star formation concentrated to the central $\sim 0\farcm6$ region \citep{1995ApJ...443L..73K}, which at a distance of 17.1 Mpc \citep{1994Natur.371..757F} corresponds to $\sim 3$ kpc. 
The starburst nature of M100 is consistent with its large number of supernovae (SNe). Since 1901 the galaxy has hosted seven SNe: SNe~1901B  and 1914A \citep{1917LicOB...9..108C,1938ApJ....88..285B}, 
1959E  \citep{1961PASP...73..175H}, 1979C \citep{1979IAUC.3348....1M}, 2006X \citep{2006IAUC.8667....1P}, 2019ehk \citep{2019TNSTR.666....1G} and 2020oi \citep{2020TNSTR..67....1F}. Whereas SN 1914A is of unknown SN type, the others have been classified as Type I (SNe 1901B and 1959E), Type Ia (SN 2006X), Type Ib (SN 2019ehk), Type Ic (SN 2020oi) and Type IIL (SN 1979C). 

In addition to optical data, there are also X-ray data for the M100 SNe. Upper X-ray limits from 1995 were derived for SNe 1901B, 1914A and 1959E by \citet{1998A&A...331..601I}, and shortly after optical detection for SN 2006X \citep{2006ATel..751....1I}. This contrasts with a wealth of X-ray detections for SN 1979C \citep[e.g.,][]{1998A&A...331..601I,2005ApJ...632..283I,2001ApJ...560..715K, 2025arXiv251103539A} as well as detections of SNe 2019ehk \citep{2020ApJ...898..166J} and 2020oi \citep{2020ApJ...903..132H}. 

The situation is similar for radio, with no data published for SNe 1901B, 1914A and 1959E, and upper limits for SNe 2006X \citep{2006ATel..954....1C,2008ATel.1393....1C,2006ATel..728....1S} and 2019ehk \citep{2020ApJ...898..166J}. However, for SN 1979C there have been published radio light curves until 2005 and even reported spatially resolved structures \citep[e.g.,][]{1986ApJ...301..790W,1991ApJ...380..161W,1992ApJ...399..672W,2000ApJ...532.1124M,2003ApJ...591..301B,2008ApJ...682.1065B,Marc09}. For SN 2020oi there are well-sampled radio and submillimeter data for the first 94 days \citep{2020ApJ...903..132H,2020ATel13448....1M,2021ApJ...918...34M}.

The common theme for the X-ray and radio detections of SNe 1979C, 2019ehk and 2020oi is circumstellar interaction, where light curves of the data have been used to derive densities and map out the structure of the circumstellar medium (CSM) in each case. In the case of SN 2006X, the absence of X-ray and radio emission may have as important implications for its CSM. This Type Ia SN (SN Ia) had clear indications of a circumstellar shell as evidenced by the time-varying narrow absorption lines \citep{2007Sci...317..924P}. The distance of the shell from the SN is uncertain and could lie further away \citep{2020ApJ...890..159L}, than probed by published radio data \citep{2008ATel.1393....1C}.

NGC 891, at a distance of 9.12 MPc \citep{2013AJ....146...86T}, is a nearby galaxy analogous to the Milky Way. As an edge-on spiral galaxy, it has been extensively studied specifically for disk and halo interactions. The galaxy hosts SN 1986J, which has been extensively studied at all wavelengths. It was first detected in radio \citep{1986IAUC}, but the explosion probably occurred in 1983 \citep{2002ApJ}. The first optical spectrum showed a narrow H$\alpha$ line \citep{1986Gunn}, and \cite{1987Rupen} put in the class of SNe that later would be coined Type IIn. Due to circumstellar interaction, it has been bright in optical \citep{2008Mili}, infrared \citep{2016Tiny}, radio \citep[e.g.,][]{2017ApJ...851....7B}, and X-rays \citep{2005Houck} for more than two decades. However, VLBI imaging also shows a spatial separation of the emission from the circumstellar interaction shell and a central component \citep{2004Sci}.

We describe the observations used for both galaxies in Section \ref{sec:obs}. For the SNe we concentrate on SNe 1979C, 1986J and 2006X. The discussions and radio modeling of SN 2006X and SN 1979C are covered in Sections \ref{sec:06x} and \ref{sec:79c}, respectively. In Section \ref{sec:86j}, we discuss the flux density differences of SN 1986J with LOFAR and the ILT and finally conclude in Section \ref{sec:conclusion}.

\section{Observations} \label{sec:obs}
\subsection{M100}

\begin{figure*}
\includegraphics[width=0.7\textwidth]{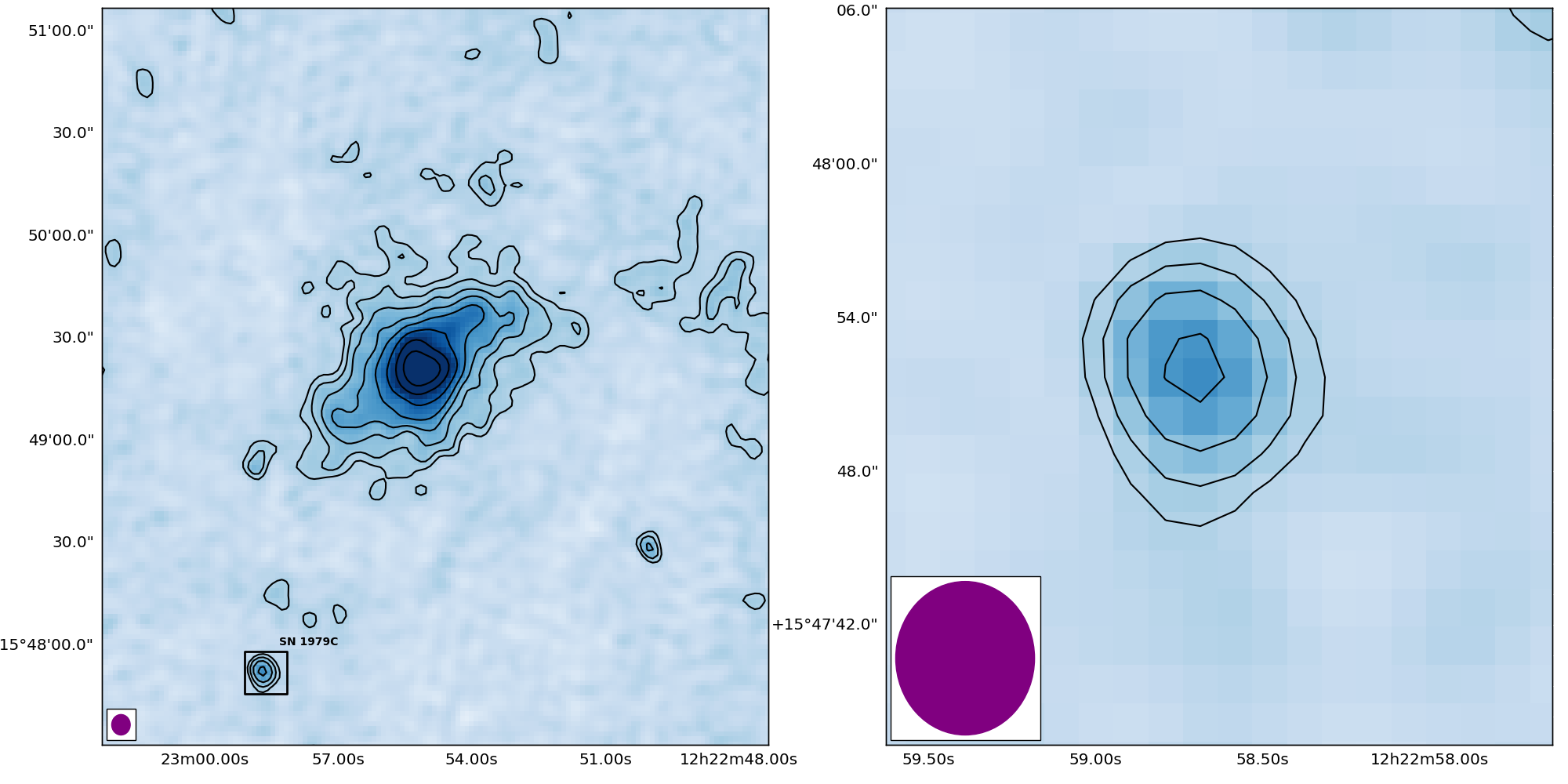}
\caption{LOFAR image of M100 at  \SI{6}{\arcsec} resolution at a central frequency of 146 MHz. Left panel shows the galaxy while the right panel is a zoom-in of SN 1979C. The purple area to the lower left in both panels marks the resolution element of the images.}
\label{fig:m100_lof}
\end{figure*}

LOFAR data of M100 at 0.146 GHz are obtained from the internal data release of widefields imaged as part of LoTSS (LOFAR Two-Metre Sky Survey; \citealt{2022A&A...659A...1S}). The image was part of pointing P185+17 and scaled by the flux-scale correction provided with the pointing. The observation from 2019-01-19 has a beam size of \SI{6}{\arcsec} with a clear detection of SN 1979C (Fig. \ref{fig:m100_lof}).

VLA data are obtained from project 19A-271 (PI:Deanne Coppejans) observed on 2019-08-29 and from project 19B-350 (PI: Assaf Horesh) observed on 2020-11-15. The calibrated measurement sets were directly available and the imaging was performed for the C-band images centered at 6.1 GHz with the CASA function \textsc{tclean}.

e-MERLIN data are obtained from two projects that observe SN 2020oi (DD9007 and CY10006). We obtain C-band images in February and August 2020 and L-band images in August 2020. For the C-band images where SN 1979C is far from the pointing center, the data sets are phase-shifted to the position of the SN. For the L-band data with a wider field of view, the \textsc{phasecenter} parameter of the CASA function \textsc{tclean} is used during the imaging. Phase self-calibration was done to improve noise for the first block of C-band observations in which a transient source (1222+1549) was bright enough. For the L-band data, eight other bright sources in the field of view were used for self-calibration. 

The observations with the upgraded Giant Metrewave Radio Telescope (uGMRT) are with band 5 (1000-1450 MHz) and band 4 (550-900 MHz). The band 4 and 5 observations were made on January 11, 2020 and January 13, 2020, respectively. The field of SN 1979C was observed in the context of SN 2020oi in M100. 3C286 was used as flux calibrator and the VLA calibrator 1120+143 was used as phase calibrator. The bandwidth for both observations was 200 MHz. However, a fraction of the bandwidth was lost due to radio frequency interference (RFI). Data were analyzed using standard CASA tasks for uGMRT. 
We carried out a Gaussian fit at the SN 1979C position to estimate the flux density.

Other VLBI and VLA data are taken from the literature, as given in Tables \ref{tab:79cfluxes} and \ref{tab:06xfluxes}.
 
\begin{table*}
\caption{SN 1979C Observations}
\scalebox{0.9}{
\begin{tabular}{lcccccc}
\tableline\tableline
Date of observation & Time after explosion  & Central Frequency &  Flux Density    & Luminosity & Instrument & Reference \cr
(UT)                     &  (years)                      &       (GHz)   &  (mJy)  & ($10^{25}$~erg~s$^{-1}$~Hz$^{-1}$)    &  &            \cr
\tableline

2001 Feb 22 & 21.89 & 1.67 & $4.28\pm0.23$ & $150\pm8$ & VLA &  \cite{2003ApJ...591..301B}    \cr
2001 Feb 22 & 21.89 & 8.46 & $2.3\pm0.4$ & $80\pm14$ & " &  "   \cr
2002 Nov 18 & 23.62 & 1.6 & $2.96\pm0.06$ & $104\pm2$ & Global VLBI &  \cite{Marc09}    \cr
2005 Feb 25 & 25.90 & 1.43 & $3.19\pm0.22$ & $112\pm8$  & " & \cite{2008ApJ...682.1065B} \cr
2005 Feb 25 & 25.90 & 4.99 & $1.68\pm0.09$ & $58.8\pm3.1$  & " & " \cr
2005 Feb 25 & 25.90 & 8.43 & $1.37\pm0.10$ & $47.9\pm3.5$  & " & " \cr
2008 Feb 19 & 28.88 & 8.4 & $0.72\pm0.028$ & $25.2\pm1.0$& VLA & This work \cr 
2019 Jan 19 & 39.79 & 0.146 & $4.6\pm0.36$ & $161\pm13$  & LOFAR & " \cr 

2019 Aug 29 & 40.40 & 6.1 & $0.36\pm0.02$ & $12.6\pm0.7$  & VLA & " \cr 

2020 Jan 11 & 40.79 & 0.65 & $1.76\pm0.09$ & $61.6\pm3.1$  & GMRT & " \cr 
2020 Jan 13 & 40.79 & 1.26 & $1.22\pm0.05$ & $42.7\pm1.7$  & " & " \cr 
2020 Jan 17 & 40.79 & 5.1 & $0.34\pm0.02$ & $11.9\pm0.7$  & e-MERLIN & " \cr 

2020 Aug 27 & 41.40 & 4.5 & $0.36\pm0.02$ & $12.5\pm0.7$ &" & " \cr 
2020 Aug 30 & 41.41 & 1.5 & $1.014\pm0.033$ & $35.5\pm1.2$  & " & " \cr 
2020 Nov 15 & 41.62 & 6.1 & $0.37\pm0.02$ & $12.9\pm0.7$ & VLA & " \cr 
\tableline
\end{tabular}
}
\tablecomments{The columns starting from left to right are as follows: Date of observation;  Time after explosion (calculated from $t_0$ = 1979.26)
; Central frequency of observation; flux densities and error from CASA IMFIT function; derived luminosity from the same; Instrument used for the observation and References. }
\label{tab:79cfluxes}
\end{table*}

\begin{table*}
\caption{SN 2006X observations}
\scalebox{0.9}{
\begin{tabular}{lcccccc}
\tableline\tableline
Date of observation & Time after explosion  & Central Frequency &  Flux Density    & Luminosity & Instrument & Reference \cr
(UT)                     &  (years)                      &       (GHz)   &  (mJy)  & ($10^{25}$~erg~s$^{-1}$~Hz$^{-1}$)    &  &       \cr
\tableline
2006 Nov 17 & 0.786 & 4.8 & $<0.126$ & $<4.41$ & VLA &\cite{2016ApJ...821..119C} \cr 
2006 Nov 20 & 0.794 & 8.4 & $<0.108$ & $<3.78$ & " & " \cr 
2008 Feb 19 & 2.043 & 8.4 & $<0.058$ & $<2.03$& " & \cite{2008ATel.1393....1C} \cr 
2019 Jan 19 & 12.96 & 0.146 & $<0.7$ & $<24.5$& LOFAR & This work  \cr 
2019 Aug 29 & 13.57 & 6.1 & $<0.02$ & $<0.70$ & VLA & "  \cr 
2020 Aug 30 & 14.57 & 1.5 & $<0.05$ & $<1.85$ & e-MERLIN & " \cr 
2020 Nov 15 & 14.78 & 6.1 & $<0.03$ & $<1.05$ & VLA & " \cr 
\tableline
\end{tabular}
}
\tablecomments{The columns starting from left to right are as follows: Date of observation;  Time after explosion (calculated from $t_0$ = 2006.094); Central frequency of observation; $3\sigma$ upper limit on the flux density; $3\sigma$ upper limit on the luminosity; Instrument and References. 
}
\label{tab:06xfluxes}
\end{table*}


\subsection{NGC 891}

\begin{figure*}
\includegraphics[width=0.7\textwidth]{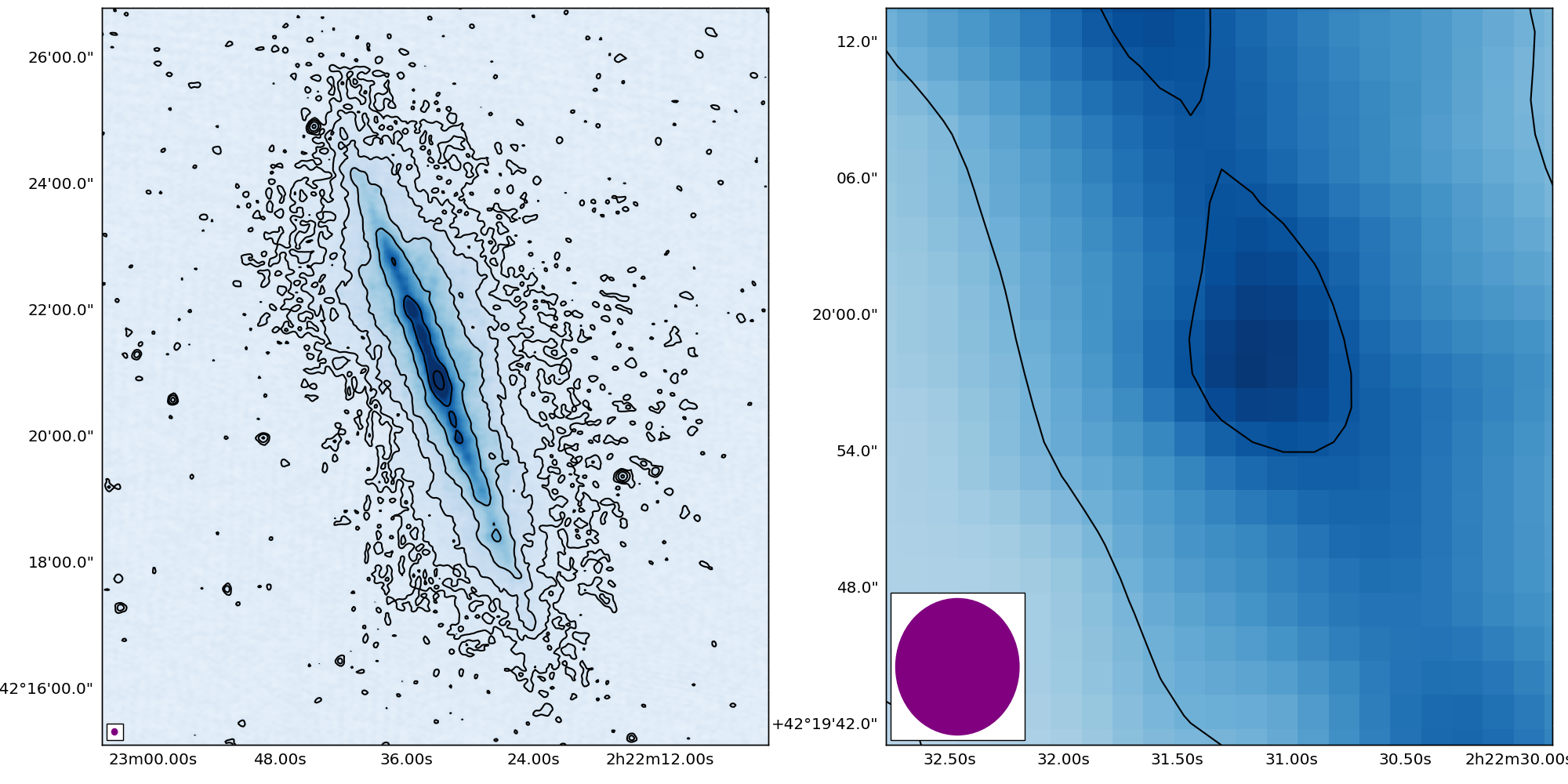}
\caption{LOFAR image of NGC 891 at  \SI{6}{\arcsec} resolution at a central frequency of 146 MHz. Left panel shows the galaxy while the right panel is a zoom-in of SN 1986J. The purple area to the lower left in the panel to the right marks the resolution element of the image.}
\label{fig:ngc891_6"}
\end{figure*}

\begin{figure*}
\includegraphics[width=0.7\textwidth]{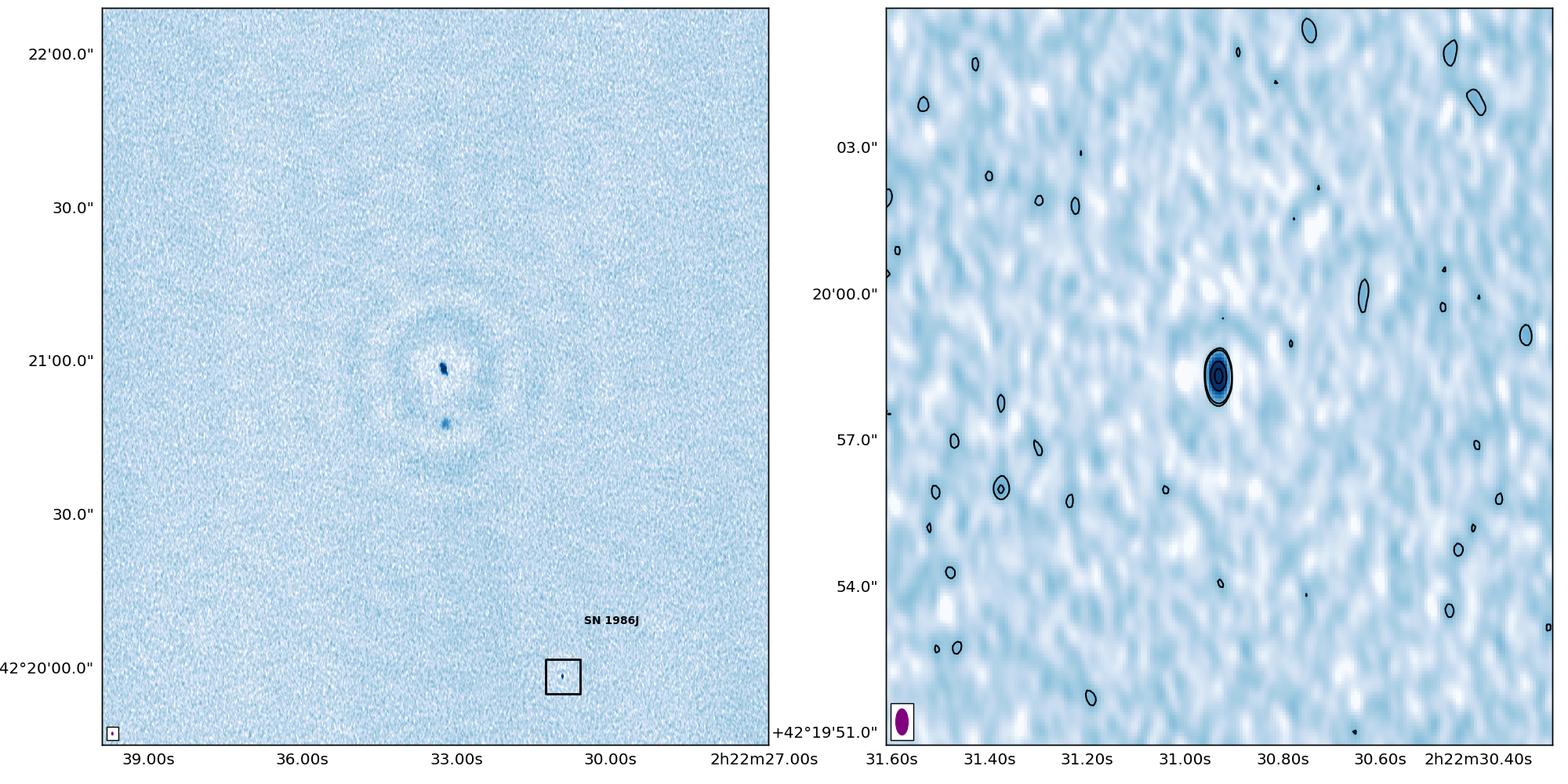}
\caption{ILT image of NGC 891 at 0\farcs54 $\times$ 0\farcs28 resolution at a central frequency of 146 MHz. Left panel shows the central part of the galaxy while the right panel is a zoom-in of SN 1986J. The purple area to the lower left in the panel to the right marks the resolution element of the image.}
\label{fig:ngc891_0.5"}
\end{figure*}


LOFAR data are obtained from project LT10\_010 (PI: T.Shimwell) observed on 8 October 2018 as part of LoTSS. The second LoTSS data release (LoTSS-DR2; \citealt{2022A&A...659A...1S}) provides a mosaic of the pointing P035+41 from this project. We used this image at \SI{6}{\arcsec} resolution to report the LoTSS flux density for SN 1986J in Table \ref{tab:86jfluxes}. In addition to this, we also process the same data set with the international stations to obtain an International LOFAR Telescope (ILT) image of NGC 891 at \SI{0.5}{\arcsec} resolution, as shown in Fig. \ref{fig:ngc891_0.5"}.

Data from the literature for LOFAR \citep{Mulcahy18}, Global VLBI \citep{2017ApJ...851....7B} and VLASS \citep{2021ApJ...923L..24S} are reported in Table \ref{tab:86jfluxes}. We use the distance from \citet{2013AJ....146...86T}, that is 9.12 Mpc. 

\begin{table*}
\caption{SN 1986J Observations}
\scalebox{0.9}{
\begin{tabular}{lcccccc}
\tableline\tableline
Date of observation & Time after explosion  & Central Frequency &  Flux Density    & Luminosity & Instrument & Reference \cr
(UT)                     &  (years)                      &       (GHz)   &  (mJy)  & ($10^{25}$~erg~s$^{-1}$~Hz$^{-1}$)    &      &        \cr
\tableline
2012 Apr 10 & 29.1$^a$ &1.10 & $1.61\pm0.10$ & $16.0\pm1.0$ & VLA & \cite{2017ApJ...851....7B} \cr
“ & 29.1 & 1.40 & $1.34\pm0.08$ & $13.3\pm0.8$ & " & " \cr
“ & 29.1 & 1.65 & $1.37\pm0.08$ & $13.6\pm0.8$ & " & " \cr
“ & 29.1 & 1.87 & $1.21\pm0.07$ & $12.0\pm0.7$ & " & " \cr
“ & 29.1 & 2.38 & $1.23\pm0.07$ & $12.2\pm0.7$ & " & " \cr
“ & 29.1 & 3.03 & $1.30\pm0.07$ & $12.9\pm0.7$ & " & " \cr
“ & 29.1 & 3.69 & $1.43\pm0.07$ & $14.2\pm0.7$ & " & " \cr
“ & 29.1 & 4.99 & $1.82\pm0.09$ & $18.1\pm0.9$ & " & " \cr
“ & 29.1 & 5.96 & $2.11\pm0.11$ & $21.0\pm1.1$ & " & " \cr
“ & 29.1 & 8.74 & $2.56\pm0.13$ & $25.5\pm1.3$ & " & " \cr
“ & 29.1 & 9.56 & $2.63\pm0.13$ & $26.2\pm1.3$ & " & " \cr
“ & 29.1 & 13.37 & $2.98\pm0.15$ & $29.7\pm1.5$ & " & " \cr
“ & 29.1 & 14.63 & $2.97\pm0.15$ & $29.6\pm1.5$ & " & " \cr
“ & 29.1 & 20.70 & $2.50\pm0.17$ & $24.9\pm1.7$ & " & " \cr
“ & 29.1 & 21.70 & $2.27\pm0.19$ & $22.6\pm1.9$ & " & " \cr
“ & 29.1 & 32.00 & $2.02\pm0.11$ & $20\pm1.1$ & " & " \cr
“ & 29.1 & 41.00 & $1.37\pm0.14$ & $13.6\pm1.4$ & " & " \cr
2013 March 31& 30.0 & 0.146  &   $5.5^b\pm0.2$& $54.7\pm2$ & LOFAR & \cite{Mulcahy18}\cr 
2014 Oct 22 & 31.6 & 5.00 & $1.62\pm0.16$ & $16.1\pm1.6$ & Global VLBI & \cite{2017ApJ...851....7B} \cr
2018 Oct 08  & 35.6   & 0.146    &  $9.66\pm0.4$  &  $96.2\pm4$  &    LoTSS  & This work \cr
"     &  35.6 & 0.146    &  $6.77\pm0.2$  & $67.4\pm2$   &   ILT  & "  \cr  
2019 Apr 15  &  36.1  & 3   & $1.3\pm0.2$ & $12.9\pm2$  & VLASS & \cite{2021ApJ...923L..24S} \cr 
\tableline
\end{tabular}
}
\tablecomments{The columns starting from left to right are as follows: Date of observation;  Time after explosion (calculated from $t_0$ = 1983.2);
Central frequency of observation; flux density; luminosity; instrument used for the observation and references. $^a$Was erroneously listed as 29.6 years in \cite{2017ApJ...851....7B}.  $^b$ \cite{Mulcahy18} remove background flux to report approximate SN flux
}
\label{tab:86jfluxes}
\end{table*}


\section{SN 2006X} \label{sec:06x}

SN 2006X was discovered on 2006 February 4.75 UT independently by Suzuki and Migliardi \citep[see][]{2006IAUC.8667....1P}, and \citet{2006CBET..393....1Q} classified it as a Type Ia SN.
\citet{2009PASJ...61..713Y}  found from spectroscopic studies that
the SN shows very high expansion velocities, especially of 
\ion{Si}{2} and \ion{S}{2}, suggesting that its spectroscopic characteristics can be explained by the delayed detonation model. In particular, \citet{2009PASJ...61..713Y} highlights that the CS15DD2 model in
\citet{1999ApJS..125..439I} appears to be compatible with the observations.. The explosion energy, $E$, and the ejecta mass, $M$, in this model are $1.44\EE{51}$~ergs 
and $1.38 \Msun$, respectively. We can approximate the ejecta structure of this model by two power laws, where the inner structure is characterized by the density profile
$\rho_i (V,t)\propto V^{-a}t^{-3}$ and the outer structure by $\rho_o (V,t) \propto V^{-n}t^{-3}$. Here, we have assumed that the ejecta are spherically symmetric and expand
homologously, that is, $V(r,t) = r/t$, where $V$ is the velocity, $r$ the radius, and $t$ the time since the explosion. The break in the density slope between these two parts of the ejecta occurs at the velocity $V_{\rm b}$, which can be calculated by integrating the density and kinetic energy profiles across the ejecta to become
\begin{equation}
	V_{\rm b}= 10\,030~\sqrt {\frac{E_{51}}{M}  \frac{\left(n-5\right) \left(5-a\right)} {\left(n-3\right)\left(3-a\right)}}~{\rm km~s}^{-1}.
\label{eq:Vb}
\end{equation}
Here $E_{51}$ is $E$ in $10^{51}$ ergs, and $M$ the total ejecta mass in solar masses \citep[see also][]{1994ApJ...420..268C,2024ApJ...976..213V}. From an inspection
of the model CS15DD2 \citep{1999ApJS..125..439I}, we find that $a=2$ and $n=12$ (together with $E_{51}=1.44$ and $M=1.38 \Msun$) give $V_{\rm b} \approx 15\,650 \kms$. These
values for $V_{\rm b}$ and $a=2$ provide a good fit to the CS15DD2 model. If we choose $n=10$, which is the preferred value for SNe Ia in e.g. \cite{2012ApJ...750..164C}, $V_{\rm b} \approx 14\,990 \kms$. The mass of the outer part of the ejecta described by power-law $n$ is 
$M_{2,{\rm out}} = (3-a) (n-a)^{-1} M$, which in our case (with 
$n=12$) is equal to $0.138 \Msun$. For $n=10$ it is $0.175 \Msun$. In the following in this section, when we compare the values derived using $n=12$ and $n=10$, we put the values for $n=10$ in parentheses.

The outer part of the ejecta will interact with the surrounding circumstellar medium (CSM). For SN 2006X this may have a complicated structure. In particular, \citet{2007Sci...317..924P} found 
time-varying narrow \ion{Na}{1} absorption features along the line
of sight to the SN, at an inferred distance of $10^{16}$ - $10^{17}$ 
cm from the SN. No narrow emission lines were detected, and radio
observations had a gap between days 18 and 287
\citep{2016ApJ...821..119C}, so any temporary radio increase could
have been missed, especially if the shell had a modest thickness \citep[see][who constructed models for radio emission in shell-like media]{2016ApJ...823..100H}. Here, we assume that the SN ejecta
expand into a constant-density medium. In this case, the interaction
can be described by a self-similar solution \citep{1982ApJ...258..790C} until the reverse shock driven into the
SN ejecta reaches the ejecta with velocity $V_{\rm ej} = V_{\rm b}$. 
The time this occurs defines time $t_{\rm b}$. For a constant
density medium
$t_{\rm b} \propto \rho_{\rm c}^{-1/3} M^{5/6} E^{-1/2}$
\citep{1982ApJ...258..790C,2024ApJ...976..213V}, where
$\rho_{\rm c}$ is the density of the medium in which the SN ejecta expand. For a helium-to-hydrogen ratio (by number) of $n_{\rm He}/n_{\rm H} = 0.1$, the density of the surrounding 
medium is $\rho_{\rm c} = 1.4$~m$_{\rm p}~n_{\rm H}$, where 
$n_{\rm H}$ is the hydrogen density in $\cm3$. $t_{\rm b}$ can
then be expressed as 
\begin{equation}
	t_{\rm b}= 86.5 
 \left(\frac{M_{1,{\rm b}}}{0.1 \Msun}\right)^{1/3} 
 \left(\frac{n_{\rm H}}{1 \cm3}\right)^{-1/3} 
 \left(\frac{V_{\rm b}}{10^4 \kms}\right)^{-1}~
 \frac{R_2}{R_1}~{\rm yrs}.
\label{eq:tb}
\end{equation}
regardless of the values for $a$ and $n$. Here $M_{1,{\rm b}}$ is the mass of the surrounding medium swept up by the forward shock at time $t_{\rm b}$, and $R_2/R_1$ is the radii ratio of the reverse and forward shocks. $M_{1,{\rm b}}$ equals 
$M_{2,{\rm out}} (M_2/M_1)^{-1}$, where $M_2/M_1$ is the mass ratio 
of the shocked ejecta and the shocked surrounding medium for times 
$t \leq t_{\rm b}$. From the similarity solutions of
\citet{1982ApJ...258..790C} we get
$M_2/M_1 \approx 1.6~(1.1)$ and
$R_2/R_1 \approx 0.869~(0.854)$ and a surrounding medium of constant density. This gives
$M_{1,{\rm b}} \approx 0.086~(0.157) \Msun$, and for
$V_{\rm b} = 15\,650~(14\,990) \kms$, we obtain $t_{\rm b} \approx 45.6~(57.0)~n_{\rm H}^{-1/3}$~ years (with $n_{\rm H}$
expressed in $\cm3$), which means that SN 2006X is still in its
ejecta-dominated phase if $n_{\rm H} \lsim 16~(32) \cm3$.

After $t\geq t_{\rm b}$, it progresses to the Sedov-Taylor stage, and the forward
shock will eventually evolve as $R_1 \propto t^{0.4}$ \citep[cf.][]{1999ApJS..120..299T}. Before then
\begin{equation}
	R_1= R_1(t_{\rm b})~(\frac{t}{t_{\rm b}})^{(n-3)/n},
\label{eq:R1}
\end{equation}
where $R_1(t_{\rm b}) \approx 2.59~(3.16)\times 10^{18}$~cm, and the forward shock velocity is 
\begin{equation}
	V_1= V_1(t_{\rm b})~(\frac{t}{t_{\rm b}})^{-3/n},
\label{eq:V1}
\end{equation}
where $V_1(t_{\rm b}) \approx 13\,510~(12\,290)\kms$.

To model the radio emission from the interaction between the ejecta and the surrounding medium, we follow what is outlined in \citet{2023ApJ...952...24H} and \citet{2024ApJ...976..213V}, that is, we assume that the fraction $\epsilon_B$ of the forward shock energy density $\rho_{\rm c} V_1^2$, goes into the magnetic field energy density, $u_B$, and the fraction $\epsilon_{\rm rel}$ goes into the relativistic electron energy density, $u_{\rm rel}$. 
We further assume that relativistic electrons have a power-law distribution of the electron energies, $n(\varepsilon) = N_0\varepsilon^{-p}$. 
Here, $\varepsilon=\gamma m_e c^2$ is the energy of the electrons and $\gamma$ is the Lorentz factor.
We refer to \cite{2024ApJ...976..213V} on how to calculate $N_0$ and $\gamma_{\rm min} m_e c^2$, which is the minimum energy of electrons that contribute to synchrotron emission.

For the postshock magnetic field energy density, $B = (8 \pi u_B)^{1/2}$, and as standard we use the value $\epsilon_B = 0.01$, which agrees with the geometric mean of $\epsilon_B \approx 0.017$ for Tycho, Kepler and G1.9+0.3 \citep{2021ApJ...917...55R}. \cite{2019ApJ...872..191S} devised an analytical approach to estimate the evolution of $\epsilon_B$, but as noted in \citep{2022Univ....8..653L}, this method does not reproduce the observed radio emission for individual SNRs very well.

For $\epsilon_{\rm rel}$, we have assumed that the geometric mean of $\sim 0.001$ for the young Type Ia SNRs Tycho, Kepler and G1.9+0.3 \citep{2021ApJ...917...55R} constitutes a minimum value for the much
younger SN 2006X, and we have tested values between $0.001-0.1$. The intensity of optically thin synchrotron emission is $\propto \nu^{-\alpha}$, where the spectral index, $\alpha$, is $\alpha = (p-1)/2$. For $p=2.8$, $\alpha = 0.7$. We have chosen this $p$ value for SN 2006X since it is steeper than for young SNRs, but somewhat shallower than for young stripped core-collapse SNe \citep{2006ApJ...651..381C,2021ApJ...917...55R}.


\begin{figure*}
\includegraphics[width=0.7\textwidth]{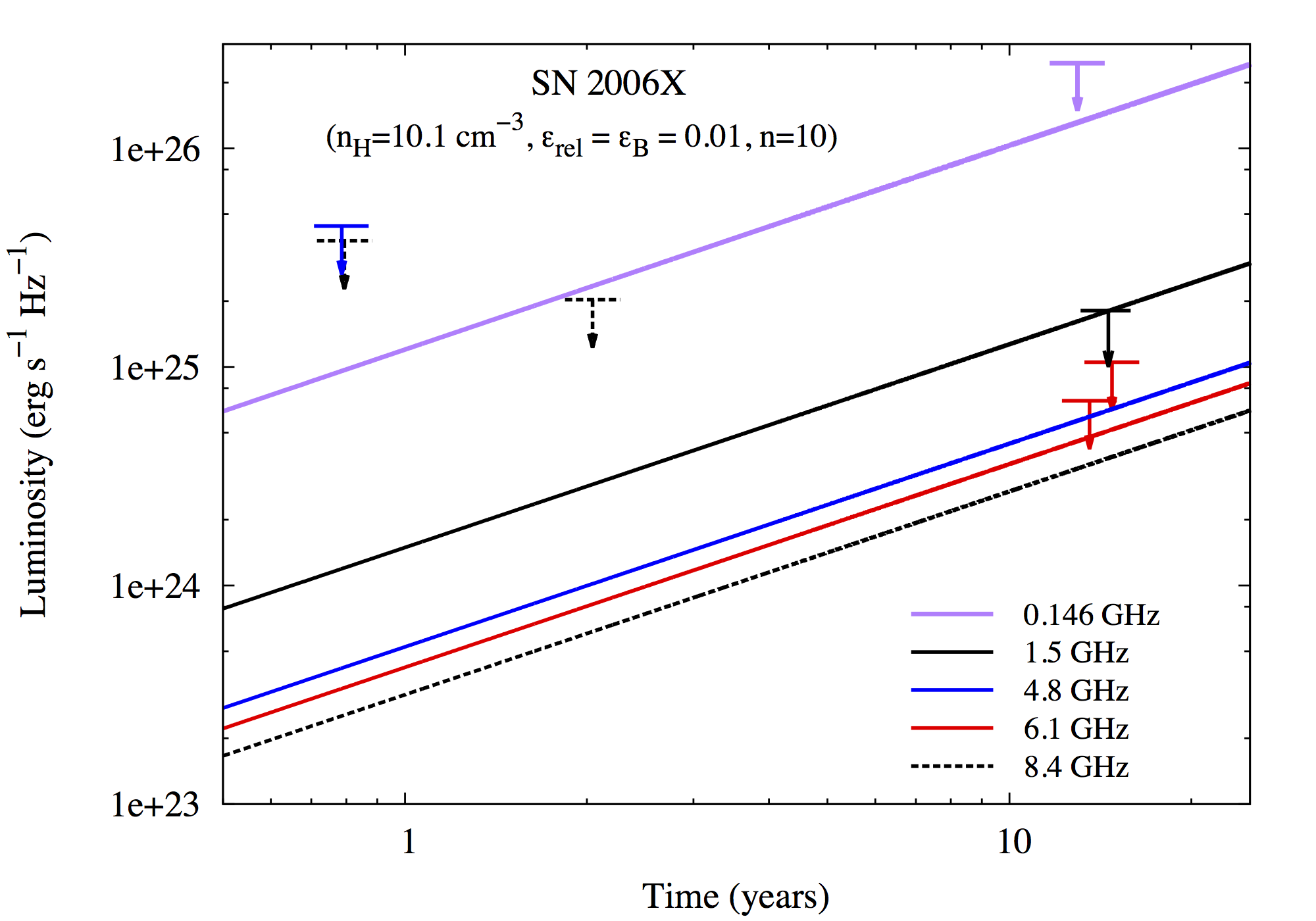}
\caption{Modeled radio light curves for SN 2006X for five frequencies (0.146, 1.5, 4.8, 6.1 and 8.4 GHz) along with all data listed in Table \ref{tab:06xfluxes}. The model parameters are $\epsilon_{\rm rel} = \epsilon_{\rm B} = 0.01$, $n=10$, $p=2.8$, and n$_{\rm H}$=10.1~cm$^{-3}$. For this value of n$_{\rm H}$ the predicted luminosity at 1.5 GHz after 14.57 years equals the observed $3\sigma$ upper limit on 2020 Aug 30, while the modeled luminosities fall below the observed $3\sigma$ upper limits of all other epochs/frequencies in Table \ref{tab:06xfluxes}. For the distance to SN 2006X we have used 17.1 Mpc. See text for further details.}
\label{fig:06_lightcurves}
\end{figure*}

\begin{figure*}
\includegraphics[width=0.7\textwidth]{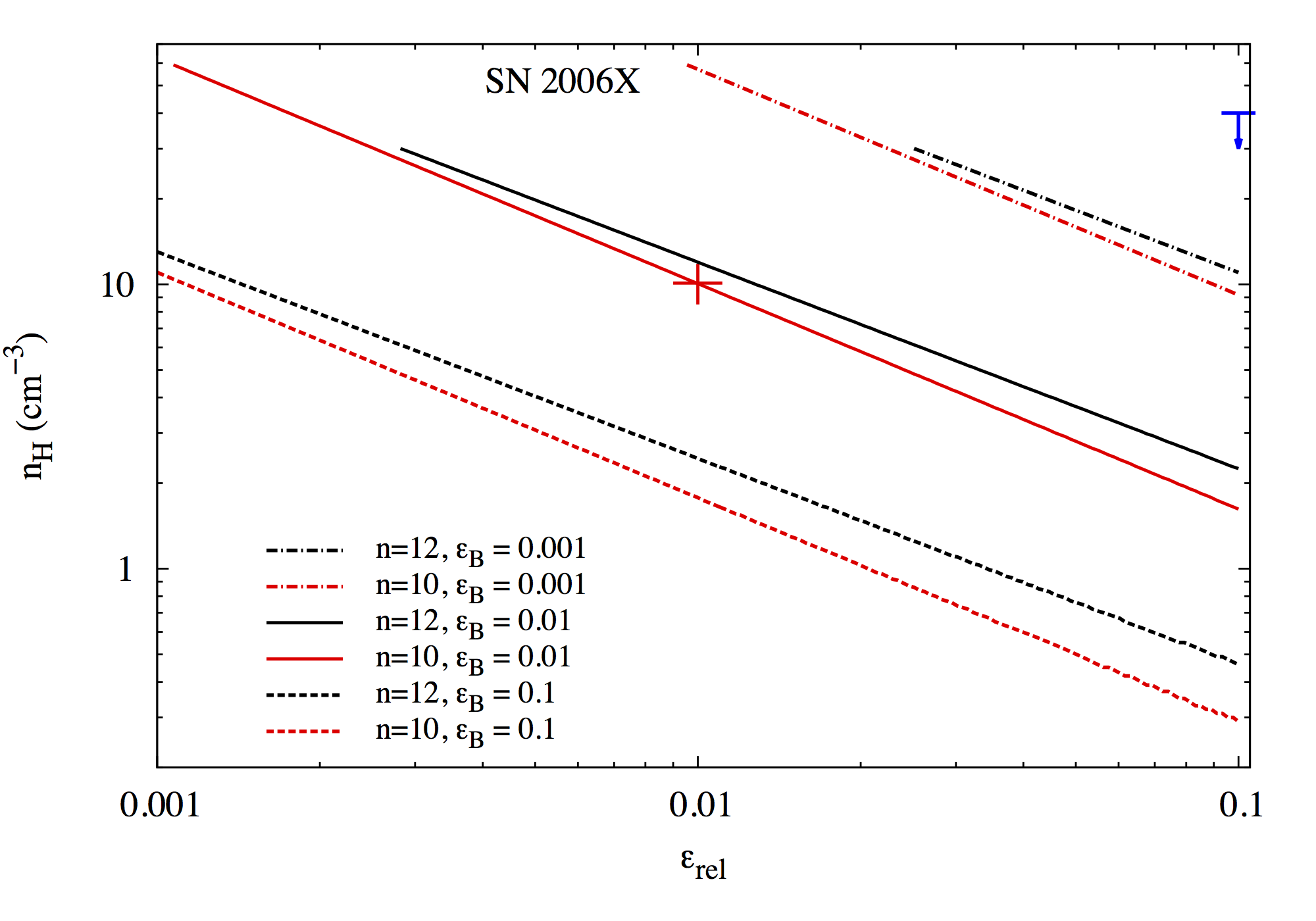}
\caption{Derived values of $n_{\rm H}$ as a function of $\epsilon_{\rm rel}$ for various models with fixed values of $\epsilon_{\rm B}$ between $0.001-0.1,$ and for $n=10$ and $n=12$. The criteria for the models are that they should produce a 1.5 GHz luminosity at 14.57 years that agrees with the observed e-MERLIN $3\sigma$ upper limit, and that 
$t_{\rm b} \geq 14.57$ years. For the three upper lines, $t_{\rm b} = 14.57$ years at their left ends, and the solution for lower values of $\epsilon_{\rm B}$ becomes much less constraining. Note that for $\epsilon_{\rm rel} = \epsilon_{\rm B} = 0.01$ and $n=10$ the value of $n_{\rm H}$ used in Figure \ref{fig:06_lightcurves} has been marked.
The upper limit at $\epsilon_{\rm B} = 0.1$ marked in blue is the derived upper limit by \cite{2016ApJ...821..119C} for $\epsilon_{\rm rel} = \epsilon_{\rm B} = 0.1$, $p=3$, $n=10.18$, and using the 4.8 GHz VLA data at 0.786 years.}
\label{fig:06_nH}
\end{figure*}

To calculate radio luminosity $L_{\nu}$ we include synchrotron self-absorption (SSA) and use the expressions in  \cite{2024ApJ...976..213V}.
The volume of synchrotron emission is assumed to be the entire volume between the reverse and forward shocks. In Figure \ref{fig:06_lightcurves} we show modeled radio light curves for the five frequencies 0.146, 1.5, 4.8, 6.1, and 8.4 GHz, together with the observed $3\sigma$ upper limits in Table \ref{tab:06xfluxes}. The model parameters are $\epsilon_{\rm rel} = \epsilon_{\rm B} = 0.01$, $n=10$, $p=2.8$, and n$_{\rm H}$=10.1~cm$^{-3}$. The parameters n$_{\rm H}$=10.1~cm$^{-3}$ were chosen so that the modeled 1.5 GHz luminosity at 14.57 years is the same as the observational $3\sigma$ upper limit. The modeled fluxes are increasing until
$t=t_{\rm b}$, so as long as $t \lesssim t_{\rm b}$, the limit on n$_{\rm H}$ can be further constrained by future observations. Figure \ref{fig:06_nH} shows the sensitivity in
n$_{\rm H}$ to $\epsilon_{\rm rel}$ and $\epsilon_{\rm B}$ for the range $0.001-0.1$ for both of these parameters. The model in Figure \ref{fig:06_lightcurves} is marked by a red plus sign in Fig.~\ref{fig:06_nH}, and the previously most sensitive limit, n$_{\rm H} =40$~cm$^{-3}$ \citep{2016ApJ...821..119C}, is shown by a blue upper limit. If we assume that $n=10$, motivated by \cite{1999ApJ...510..379M}, and n$_{\rm H} =1$~cm$^{-3}$ this places limits on $\epsilon_{\rm rel}$ and $\epsilon_{\rm B}$; if $\epsilon_{\rm rel} = 0.1$, then $\epsilon_{\rm B} \lesssim 0.021$, or if $\epsilon_{\rm B} = 0.1$, then $\epsilon_{\rm rel} \lesssim 0.019$. In Figure \ref{fig:06_nH} solution curves end on the left, well within the plot. In these cases, $t=t_{\rm b}$ at these densities, and higher densities cannot be well probed, since radio emission likely decreases with time after $t_{\rm b}$, as discussed by, e.g., \cite{2019ApJ...872..191S}.

\section{SN 1979C} \label{sec:79c}
SN 1979C was first detected at 4.86 GHz at $\sim 1$ year, and was then extensively monitored mainly at 1.49 GHz and 4.86 GHz for the first $\sim 19.5$ years \citep{1986ApJ...301..790W,1991ApJ...380..161W,2000ApJ...532.1124M}. The emission became optically thin first at the highest frequencies and at 1.49 GHz after $\sim 3-4$ years. It then declined and could be modeled relatively well with a model similar to that described in Section \ref{sec:06x} with free-free absorption included \citep{1986ApJ...301..790W}, and where the circumstellar medium is in the form of a stellar wind instead of a constant-density medium. It was realized by \cite{1991ApJ...380..161W,1992ApJ...399..672W} that the wind must have had periodic variations of $\sim 20$\% with a period of $\sim 1575$ days, and \cite{1996MNRAS.282.1018S} successfully modeled these variations for the first $\sim12$ years as a result of the supernova ejecta interacting with a steady progenitor wind modulated by a binary companion. However, subsequent data up to $\sim19.5$ years \citep{2000ApJ...532.1124M} show a radio brightening, and these authors estimate that this could be due to a density enhancement of $\sim 34$\% compared to a $\rho \propto R^{-2}$ wind. Rebrightening appears to have peaked at $\sim 16-17$ years. At 19.5 years, the 1.49 GHz flux is $5.14\pm0.41$ mJy \citep{2000ApJ...532.1124M}, which can be compared with the data for 21.89 years in Table \ref{tab:79cfluxes}. An extrapolation of those fluxes to 1.49 GHz gives a flux of $\sim 4.5$ mJy at 21.89 years, which is close to the flux at 19.5 years, but $\sim 4.5$ times larger than the 1.5 GHz flux at 41.41 years. If we extrapolate in time the results of \cite{2000ApJ...532.1124M}, only a factor of $\sim 2$ flux decrease  is expected during this time interval for a $\rho \propto R^{-2}$ wind and a constant density ejecta parameter $n$. This could indicate a dramatic change in the wind density encountered by the forward shock. It could also be the result of the reverse shock interacting with supernova ejecta with a shallower density profile.

VLBI studies of SN 1979C between $4-20$ years show that the radius of the radio-emitting plasma region increases with time as $R_{\rm radio} \propto t^m$, with $m=0.95\pm0.03$ \citep{2003ApJ...591..301B} or $m=0.91\pm0.09$ \citep{Marc09}, with a possible increased retardation (i.e., lower $m$) after $\sim 17$ years \citep{2003ApJ...591..301B}. 
If one assumes that the supernova ejecta expand into a steady spherically symmetric pre-supernova stellar wind characterized by a mass loss rate $\dot M_w$ and a wind velocity $v_w$, the wind density is described by $\rho_w=\dot M_w/(4\pi v_w R^2)$, and the self-similar solution for the expansion of the circumstellar shock \citep{1982ApJ...258..790C} can be written as
\begin{equation}
	R_1= R_1(t_{\rm b})~(\frac{t}{t_{\rm b}})^{(n-3)/(n-2)}.
\label{eq:R1w}
\end{equation}
Equation \ref{eq:R1w}, together with VLBI observations, indicates that $n \gtrsim 8$.
The absolute value of the radius derived from the VLBI data depends on the geometry of the radio-emitting region. \cite{Marc09} assumed a 30\% wide shell of emission, with an outer radius $R_{\rm radio}$, which they estimate to be 2.27 mas at $t=20.12$ years. With a distance to the supernova of 17.1 Mpc, this means $R_{\rm radio} \approx 5.81\times10^{17}$ cm at 20.12 years.
We therefore adopt $R_1({\rm 20.12~years}) \approx 5.81\times10^{17}$ cm, corresponding to an average velocity of $\approx 9\,150$~km~s$^{-1}$ during 20.12 years.
The velocity of the forward shock is 
\begin{equation}
	V_1= V_1(t_{\rm b})~(\frac{t}{t_{\rm b}})^{-1/(n-2)},
\label{eq:V1w}
\end{equation}
so, at 20.12 years it is $[(n-3)/(n-2)$]~9\,150~km~s$^{-1}$, i.e.,
$V_1({\rm 20.12~years}) \gtrsim 7\,630$~km~s$^{-1}$ since $n\geq 8$, and $t_{\rm b} \geq 20.12$~years (see below). 

The maximum velocity of the supernova ejecta is $V_{ej} = (R_1 /t) (R_2 / R_1)$, where $R_2 / R_1$ is from similarity solutions \citep{1982ApJ...258..790C, 1994ApJ...420..268C}. 
This means $V_{ej} \gtrsim (R_1 /t) (R_2 / R_1)$, which for 
$8 \leq n < \infty$ gives $7\,050 \lesssim V_{ej}/({\rm km~s}^{-1}) \lesssim 7\,690$ at 20.12~years. 
This fits the picture that broad optical emission lines emerge from the fast ejecta photoionized by X-rays, coming mainly from the reverse shock moving into the ejecta \citep{2005ApJ...632..283I}. The width of the lines indicate ejecta velocities $\gtrsim 6\,700$ km~s$^{-1}$ at $t=29.0$ years and perhaps slightly in excess of $\sim 7\,000$ km~s$^{-1}$ at $t=14$ years \citep{2009ApJ...692..839M}.  

We have assumed a similar general structure for the ejecta of SN 1979C as we did for SN 2006X, that is we use Equation \ref{eq:Vb}. As discussed in \citet{2024A_and_A...691A.171M} for SN 1993J, a rapid downturn of the radio flux is expected to occur at $t = t_{\rm b}$, as well as a faster retardation of the forward shock. At the same time, the X-ray emission from the reverse shock should fall, as it did for SN 1993J \citep{2009ApJ...699..388C}, and thus also the optical line emission (although there is still some ionizing radiation from the forward shock). One may expect that this would occur when the reverse shock has traversed the hydrogen-rich part of the supernova envelope, as is indicated to have been initiated by the disappearance of H$\alpha$ emission until $t=29.0$ years \citep{2009ApJ...692..839M}. There is also the possibility that rapidly decaying radio (and X-ray) emission could be due to a significant drop in circumstellar density encountered by the forward shock, but guided by simulations for SN 1993J \citep{2019ApJ...875...17K}, this alone does not provide fast enough decay of radio and X-rays for that supernova.

However, no observed rapid fall in X-rays is observed for SN 1979C between $16-28$ years \citep{2005ApJ...632..283I,2011NewA...16..187P}
and radio observations 
between 2001-2005 indicated only a modest fall \citep{2008ApJ...682.1065B}, although this may have been a continuation of the seemingly achromatic modulations of the radio emission between $4-20$ years \citep{2000ApJ...532.1124M,2003ApJ...591..301B}. 
To take into account the vanishing emission of H$\alpha$ by $t=29.0$ years \citep{2009ApJ...692..839M}, and to agree with observations at other wavelengths, we initially assume
that $t_{\rm b} = 40$ years.  

From an extrapolation of 20.12 years,
we estimate that $6\,280 \lesssim V_{\rm b}/({\rm km~s}^{-1}) \lesssim 7\,690$ for $8\leq n < \infty$ at $t= t_{\rm b} = 40$ years. For $\rho_w \propto R^{-2}$, $t_{\rm b} \propto (\dot M_w/v_w)^{-1} M^{3/2} E^{-1/2}$.
\begin{figure*}
\includegraphics[width=0.7\textwidth]{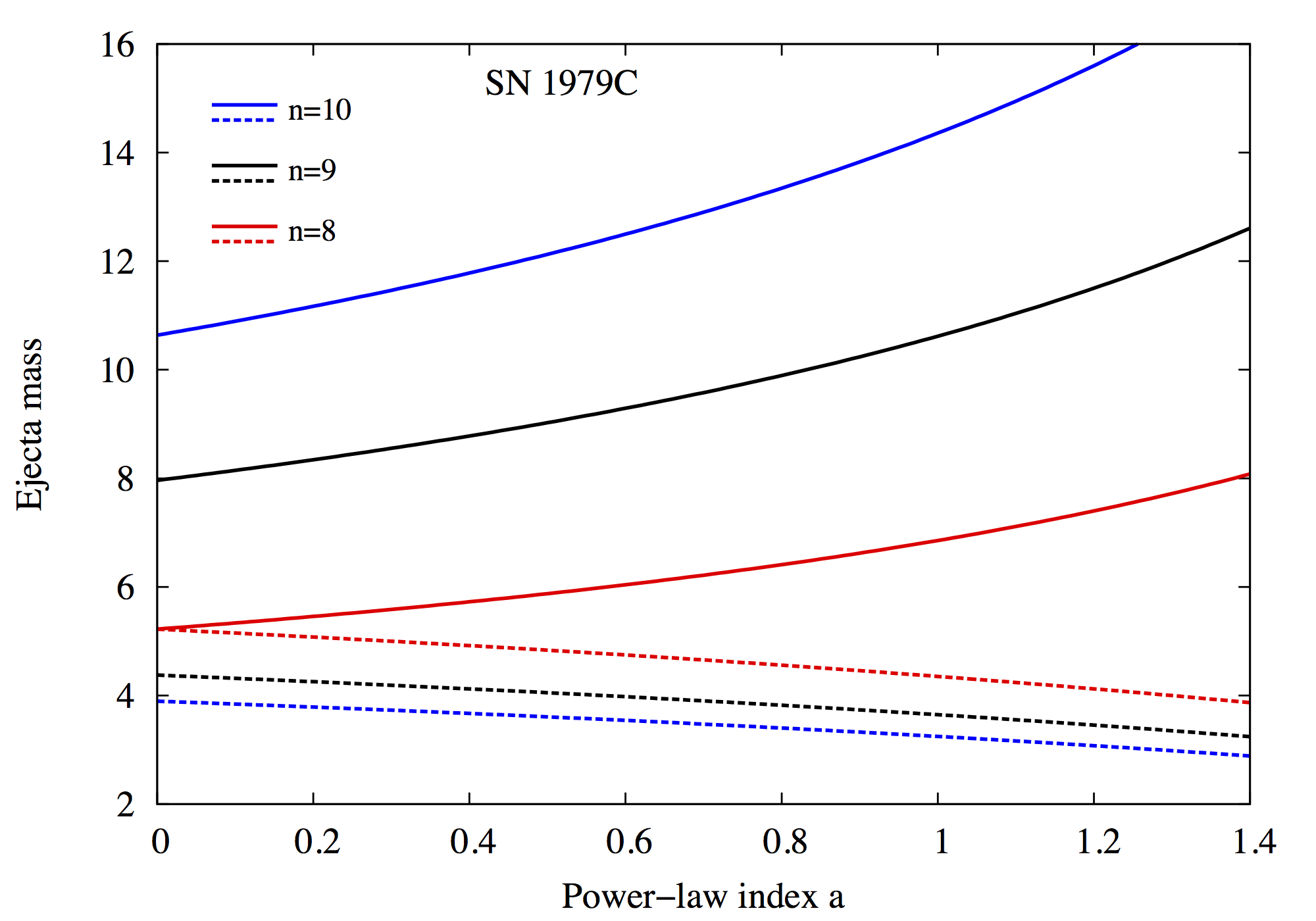}
\caption{Ejecta mass of SN 1979C as functions of the power-law index $a$ for the inner part of the ejecta, $\rho \propto V^a$. Solutions are calculated for three values of $n$, defined as $\rho \propto V^n$ for the outer part of the ejecta. Solid lines are from Equation \ref{eq:m_tot} and dashed from Equation \ref{eq:m_tot2}. When lines of the same color cross, there is a solution, which is only the case for $n=8$ at $M\approx 5.2 \Msun$. (See text for details.)}
\label{fig:79C_M}
\end{figure*}
The ejecta mass can be estimated from the amount of swept-up circumstellar mass at
$t_{\rm b}$, $M_{1,40}$, which is
\begin{equation}
	M_{1,40}= 3.27
 \left(\frac{\dot M_w}{10^{-4} \Msun~{\rm year}^{-1}}\right)
 \left(\frac{v_w}{10~\kms}\right)^{-1} 
 \left(\frac{t_{\rm b}}{40~{\rm years}}\right)^{(n-3)/n-2)} 
~\Msun,
\label{eq:m1_128}
\end{equation}
and the total ejecta mass is 
\begin{equation}
	M= \left(\frac{n-a}{3-a}\right)
 \left(\frac{M_2}{M_1}\right)
 M_{1,40}. 
\label{eq:m_tot}
\end{equation}
This should be equal to the ejecta mass from Equation \ref{eq:Vb}
\begin{equation}
	M= E_{51} 
 \left(\frac{(n-3)(3-a)}{(n-5)(5-a)}\right)
 \left(\frac{V_{\rm b}}{10\,030~\kms}\right)^{-2} ~\Msun.
\label{eq:m_tot2}
\end{equation}
 
In Figure~\ref{fig:79C_M} we show the ejecta mass $M$ as a function of parameter $a$ from Equations \ref{eq:m_tot} and \ref{eq:m_tot2} for the $n$-values 8, 9 and 10. Fixed parameters are $E_{51} = 2.05$, and $\dot M_w = 5\times10^{-5}~(v_w/10~\kms)$ $\Msun~{\rm year}^{-1}$. 
The solutions shown in Figure~\ref{fig:79C_M} are obtained when lines of the same color cross one another, which means that there is no solution for
$n \geq 9$. However, for $n=8$ we get a solution for $a\approx 0$, and $M \approx 5.2 \Msun$. We therefore choose $n=8$ and $a=0$ as our preferred values.
For this solution $M_{1,40} \approx 1.63\Msun$, and the mass of the ejecta swept by the reverse shock $M_{2,40} \approx 2.13\Msun$. This is similar to $E_{51} = 1-2$ and $M \sim 6 \Msun$ estimated by \citet{1992SvAL...18...43B} from multigroup radiation-hydrodynamic modeling for the first months of SN 1979C, with somewhat better fits for $E_{51} = 2$ than for $E_{51} = 1$. If we assume a compact object of $\sim 1.4 \Msun$, the helium core mass after hydrogen burning in our estimate is $\sim 4.5 \Msun$, which points to a zero age mass of $\lsim 15 \Msun$ \citep{1995ApJS..101..181W}, and probably close to $13 \Msun$ \citep{2013ApJ...764...21C}, although it is not clear that the oxygen mass of $\sim 0.3 \Msun$ in a $13 \Msun$ star \citep{1995ApJS..101..181W} is high enough to produce the observed emission line fluxes in SN 1979C. In this context, we note that our estimate of $13 \Msun$ is lower than $17-18 \Msun$ discussed by \cite{1999PASP..111..313V} and \cite{2000ApJ...532.1124M}. 

A completely different interpretation of the X-ray emission from SN 1979C was presented by \cite{2011NewA...16..187P}. These authors suggested that the steady X-ray luminosity could be evidence for a stellar-mass ($5-10 \Msun$) black hole accreting matter from supernova fallback or a binary companion, or that it could signal emission from a central pulsar wind nebula. The idea of a central compact object lends some support from a possible flattening of the radio spectrum around the year 2005 \citep{2008ApJ...682.1065B}; the spectral index is $\alpha = 0.63\pm0.03$ for $t \sim 18$ years \citep{2000ApJ...532.1124M}, flattening to $\alpha = 0.38\pm0.15$ for $t \sim 22$ years and $\alpha = 0.49\pm0.09$ for $t \sim 26$ years \citep{2008ApJ...682.1065B}. We do not see such flattening for the 2019 and 2020 data. Instead, the spectrum at this epoch is steeper, and the spectral index is closer to that estimated by \cite[][see below]{2000ApJ...532.1124M}.

To calculate the radio emission, we use the same model as for SN 2006X. The only difference is that we are not using a constant density for the circumstellar medium. Assuming $n_{\rm He}/n_{\rm H} = 0.1$, so that the density of the surrounding medium is $\rho_{\rm w} = 1.4$~m$_{\rm p}~n_{\rm H,w}$, we can write
\begin{equation}
	n_{\rm H,w} = 215~ 
\left(\frac{\dot M}{10^{-4} \Msun~{\rm year}^{-1}}\right)
\left(\frac{v_{\rm w}}{10~\kms}\right)^{-1} 
\left(\frac{R}{10^{18}~{\rm cm}}\right)^{-2} 
~{\rm cm}^{-3}.
\label{eq:nH}
\end{equation}
For our preferred values at 20.12 years, that is, $\dot M_w = 5\times10^{-5}~(v_w/10~\kms)$ $\Msun~{\rm year}^{-1}$ and $R_1 = 5.81\times10^{17}$~cm, $n_{\rm H,w}(R_1) \approx 320$~cm$^{-3}$. The value of $\dot M/v_w$ is about a factor of two lower than that found from time-dependent photoionization calculations by \cite{1988LF} to estimate the free-free optical depth, $\tau_{\nu,{\rm ff}}$, through the photoionized wind external to $R_1$. Because $\tau_{\nu,{\rm ff}}\propto (\dot M/v_w)^2 T_w^{-3/2}$, where $T_w$ is the temperature of the ionized wind, a lower temperature directly translates into a lower estimated value for $\dot M/v_w$. \cite{1988LF} found that $T_w \sim 3\times10^4$~K at the time the 5 GHz emission becomes optically thin, and this temperature was later used in, e.g., \cite{2000ApJ...532.1124M}. However, a lower temperature is possible, especially if the wind would be clumpy, so our preferred value of $\dot M/v_w$ seems reasonable and is similar to the estimate of \cite{1982ApJ...259..302C}, who also obtained $\dot M_w = 5\times10^{-5}~(v_w/10~\kms)$ $\Msun~{\rm year}^{-1}$. Our scenario indicates that the wind outside $R_1$ at $t_b = 40$~years could be as massive as $\sim 5 \Msun$. After $t_b$ the forward shock will continue to move into this, albeit with a continuously lower $m-$value until the expansion settles in a Sedov-Taylor solution with $R_1 \propto t^{2/3}$ for $\rho_w \propto R^{-2}$ \citep[cf.][]{1999ApJS..120..299T}. At that point the radio emission would start to decay even faster.

Although the decaying H$\alpha$ emission indicates that we may be close to $t = t_{\rm b}$, $t_{\rm b}$ could be greater than 40 years. 
Because $t_{\rm b} \propto (\dot M_w/v_w)^{-1} M^{3/2} E^{-1/2}$, one can increase $t_{\rm b}$ by increasing $M$ 
or decreasing $E$, or both. A lower $\dot M_w/v_w$ seems less likely. As an example, we can keep $n=8$ and $a=0$ 
and increase $t_{\rm b}$ to 47 years so that the mass of the ejecta, according to Equation \ref{eq:m_tot}, becomes $6 \Msun$, that is the mass studied by \citet{1992SvAL...18...43B}. This could also favor a somewhat more massive progenitor. However, Equation \ref{eq:m_tot2} then gives $E_{51}\approx 2.22$. Thus, there is no unique solution, but stretching $E$ from an already high value to an even higher value could be problematic. Continued monitoring of SN 1979C should eventually constrain $t_{\rm b}$ better.

\begin{figure*}
\includegraphics[width=0.7\textwidth]{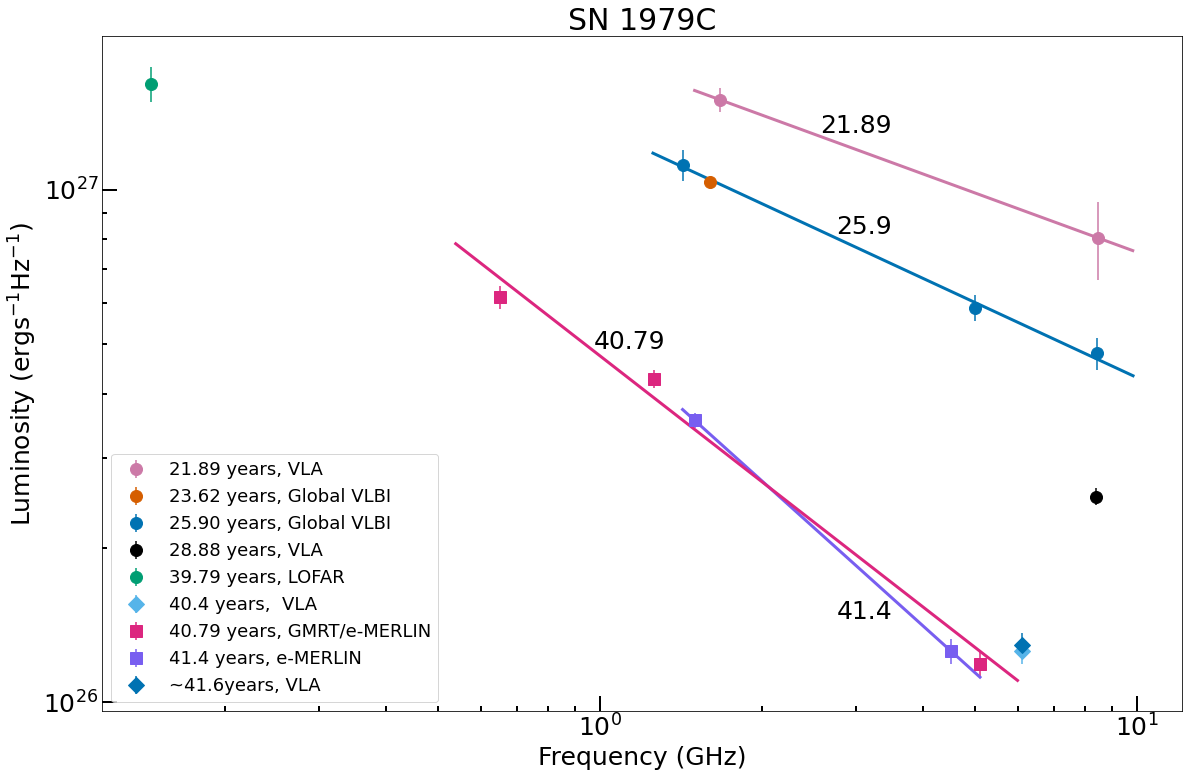}
\caption{Power-law fits to observed spectra for SN 1979C between 21.89 years and 41.4 years. The power-law index $\alpha$, defined as luminosity $L_{\nu} \propto \nu^{-\alpha}$ are, in order of increasing time: $0.38$, $0.49\pm0.03$, $0.81\pm0.1$, and $0.94$. Lines and data points of the same color are for the same age as defined in the figure. Note the spectral steepening with age, and a possible spectral break between $1-2$ GHz for $\sim 41$ years.} 
\label{fig:79c_hydro_pl}      
\end{figure*}

\begin{figure*}
\includegraphics[width=0.7\textwidth]{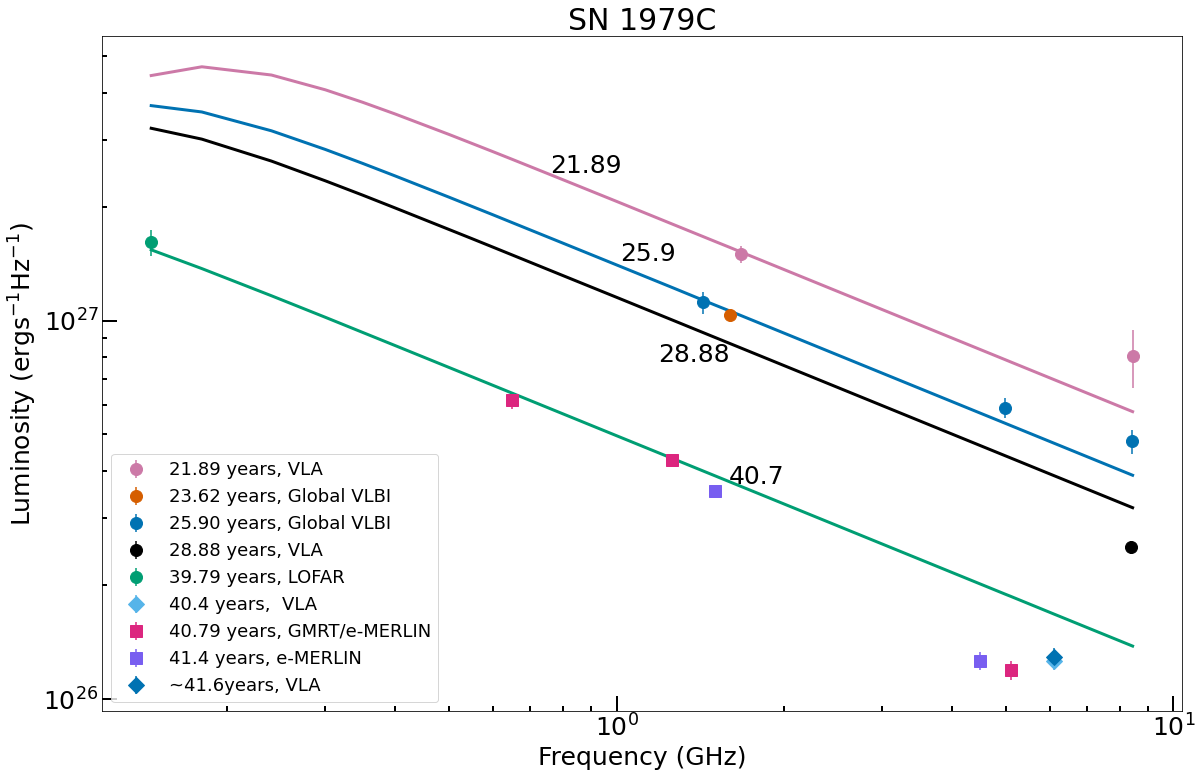}
\caption{Data points show observed luminosity at various radio frequencies and epochs for SN 1979C. The distance to the supernova has been assumed to be 17.1 Mpc. Data are from Table \ref{tab:79cfluxes}. Results from our model with $n=8$ are shown for the epochs 21.89, 25.9, 28.88 and 40.7 years. The model assumes a relativistic particle spectral index of $p=2.2$. The model undershoots at high frequencies at 21.89 years and overshoots at 40.7 years. Note the spectral turnover at low frequencies for 21.89 years due to synchrotron self-absorption. See text for further details.}
\label{fig:79c_hydro}    
\end{figure*}

Figure~\ref{fig:79c_hydro_pl} shows the radio data for SN 1979C and Figure~\ref{fig:79c_hydro} our modeling with $n=8$ and $a=0$. In this model, we have used $p=2.2$, so that $\alpha = 0.6$. This is close to what was found for for $t \sim 18$ years \citep{2000ApJ...532.1124M}, and fits the combined data for 2019-2020 ($t=40.7$ years) for the lowest frequencies\footnote{Because we have chosen $t_{\rm b}=40$ years, stretching model predictions to 40.7 years is technically outside the time range for the model, but the the effect is small.}. Although the fit is mainly for these data, we also show what the model predicts for $t=21.89$, $t=25.9$ years, and $t=28.88$ years. In general, the model results fit the data well below $1.7$ GHz at all epochs in Figure \ref{fig:79c_hydro}, but the difference is striking at frequencies higher than $1.7$ GHz, where the observations show a large spread in the spectral shape for the different epochs. 
This is clearly displayed in Figure~\ref{fig:79c_hydro_pl}. In particular, the reported spectral flattening for the first two epochs \citep{2008ApJ...682.1065B} changes to a steepening for $\sim 41$ years with a spectral index of $\approx -0.94$ at 41.6 years for frequencies above 1.5 GHz. The rapid decay of flux between 22 and 42 years decreases roughly as $F_{\nu} \propto t^{-2.1}$. This appears to be well described by our $n\approx8$ model where the flux decreases as $F_{\nu} \propto t^{-2.3}$. The interpretation that the spectral flattening could signal a central object is therefore contradicted by the combined data from 2019--2020. 

In the model in Figure \ref{fig:79c_hydro}, we have used $\epsilon_{\rm rel} = 0.005$ and $\epsilon_{\rm B} = 0.01$, inspired by values derived for young SNRs \citep{2021ApJ...917...55R}. So, the value of $B$ is $5.2$ mG at 40 years. Note the effect of synchrotron self-absorption at the lowest frequencies in the first
epochs in Figure~\ref{fig:79c_hydro}. It is small and does not play any role for $\gtrsim 40$ years. Neither is free-free absorption important for our model. It affects the flux by $\lsim 1\%$ even at the lowest frequency in the first epoch shown in Figure \ref{fig:79c_hydro}. For the last epoch shown in the figure, the
spectral break at $\approx 1.5$ GHz is therefore not due to any absorption process. However, a steepening from $\alpha \approx 0.6$ (cf. Fig.~\ref{fig:79c_hydro}) to $\alpha \approx 0.94$ 
at $\approx 1.5$ GHz (cf. Fig.~\ref{fig:79c_hydro_pl}) could be due to synchrotron cooling. The time scale for this is $\tau_{\rm synch} = 6\times10^{11} \nu_{\rm break}^{-1/2} B^{-3/2}$~s \citep{1970ranp.book.....P,2014ApJ...780...50B}. For our model, with $\nu_{\rm break} = 1.5$ GHz, $\tau_{\rm synch} \sim 1300$ years, which is too long, unless 
the value of $B$ is underestimated in our model by a factor of $\sim 10$. This means that $\epsilon_{\rm B}$ would have to be close to its formal maximum unity value (corresponding
to $B \approx 52$ mG), and that $\epsilon_{\rm rel}$ must be reduced from 0.005 to $\sim 1.3\times10^{-4}$ for the flux to remain at the same level as in Figure \ref{fig:79c_hydro}. In this context, we point out that this is similar to what \cite{2004ApJ...604L..97C} suggested for a spectral break at $\approx 4$ GHz on day 3\,200 
for SN 1993J. These authors argued that the ratio $u_{\rm rel}/u_{\rm B}$ should be in the range $8.5\times10^{-6} - 5.0\times10^{-4}$ for SN 1993J at this epoch, which brackets our value $\sim 1.3\times10^{-4}$ to obtain a cooling break at 1.5 GHz for SN 1979C at 40 years. Therefore, complete dominance of $u_{\rm B}$ over $u_{\rm rel}$ 
seems to be required for the spectral breaks to appear in the two supernovae. 

If we extrapolate our model Figure~\ref{fig:79c_hydro} back, to $t=5.5$ years there is a significant difference with the data. The model gives a 1.43 GHz luminosity of $3.3\times10^{28}$~erg~s$^{-1}$~Hz$^{-1}$, but the observed luminosity is $\sim 10$ times lower. The simplest solution to this, which at the same time would agree with the possible increased retardation derived from the VLBI measurements after $\sim 17$ years, is that there is one more break in the density profile of the ejecta to a higher value $n$. (In reality, there could be a continuous flattening of the density profile toward lower ejecta velocities.) If ejecta with $n=14$ are
attached to the ejecta with $n=8$, and that $R_1 = 5.05\times10^{17}$ cm at 17 years when we switch from $R_1 \propto t^{11/12}$ to $R_1 \propto t^{5/6}$, with a corresponding change in $V_1$, then the model gives $3.3\times10^{27}$~erg~s$^{-1}$~Hz$^{-1}$ at 5.5 years and $1.3\times10^{27}$~erg~s$^{-1}$~Hz$^{-1}$ at 17 years, which is close to the observed values given the achromatic undulations. Here we have used the same epsilon parameters as for the pure $n=8$ model. Although it takes at least one dynamical time scale to switch between models with different values of $n$, this example makes it plausible that the ejecta profile is steep ($n\sim 14$) until around 17 years and then becomes more like $n=8$. This is just before the observed change in H$\alpha$. Further flattening of the density profile of the ejecta is expected after $\sim 40$ years. Continued monitoring and detailed hydrodynamic simulations are needed to test this. We note that a double break density profile will reduce the estimated ejecta mass and kinetic energy compared to the pure $n=8$ model with $E_{51}=2.05$ and $M=5.2 \Msun$. For $t_{\rm b} = 40$ years and 17 years as a break between $n=8$ and $n=14$, $E_{51}=1.56$ and $M=4.65 \Msun$. In particular, the value for $E_{51}$ is probably more in line with the expected supernova kinetic energies.

At the time our work was being completed, late X-ray data between 26.84$-$40.90 years for SN 1979C were presented in \cite{2025arXiv251103539A}. The authors find that the data between 31.9$-$40.9 years are best explained with a soft component with a temperature of $kT \sim 0.7-1.1$ keV and a harder component with $kT \sim 1-3$ keV, which may also be a power-law rater than thermal. There is a trend of decaying
flux from the soft component with time, while the harder component could remain nearly constant. In our $n=8$ model, with pure helium in the reverse shock, the reverse and forward shock temperatures at 40.7 years are $1.8\times10^7$ K and $6.3\times10^8$ K, respectively, corresponding to 1.58 keV and 54 keV. The electron temperatures are lower, and for Coulomb heating alone, the electron temperatures  at 40.7 years are $\sim 86\%$ and $\sim 7\%$ of the shock temperatures so that they correspond to $\sim 1.37$ keV and $\sim 3.8$ keV, respectively, following the method of \cite{2019ApJ...875...17K}. If we allow for a fifty-fifty composition of hydrogen and helium by number at the reverse shock, we obtain a shock temperature for the electrons which is $\sim 1.14$ keV. We find this very similar to the values derived from observations, especially since the average temperature of the emitting shocked ejecta is expected to be lower than at the reverse shock for a $\rho \propto R^{-2}$ circumstellar medium \citep{1982ApJ...258..790C}. However, more detailed modeling is required for a full comparison between data and our model. This is beyond the scope of the present study. 



\section{SN 1986J} \label{sec:86j}
\begin{figure*}
\includegraphics[width=0.7\textwidth]{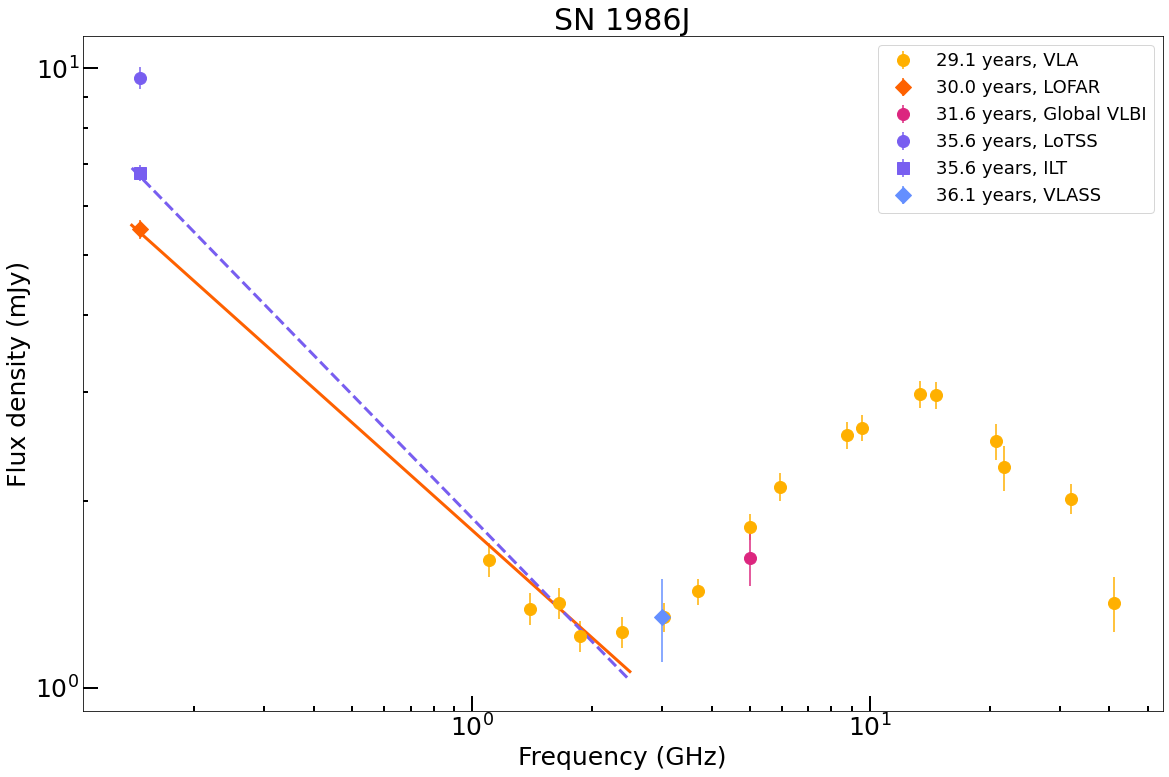}
\caption{Observed fluxes of SN 1986J between $29.1-36.1$ years of age (cf. Table~\ref{tab:86jfluxes}), assuming the supernova exploded on 1983.2. Power-law fits $F_{\nu} \propto \nu^{-\alpha}$ have been drawn for the low-frequency part of the spectrum, and indicate $\alpha = 0.58\pm0.03$ (with LOFAR) and $\alpha = 0.66\pm0.03$ (with ILT). Original fit from only VLA data was estimated to be $\alpha = 0.63\pm0.03$ \citep{2017ApJ...851....7B}.}
\label{fig:86jflux}
\end{figure*}
In Figure~\ref{fig:86jflux} we show the observed radio spectra of SN 1986J between $29.1-36.1$ years after the explosion (assumed to have occurred at $t_0 = 1983.2$). Data from $t=29.1$ years are from \citet{2017ApJ...851....7B}, and end in the low-frequency part of the spectrum at 1.10 GHz (cf. Table~\ref{tab:86jfluxes}). We extended this to 0.146 GHz with a near-contemporaneous LOFAR observation by \citet{Mulcahy18}, as well as more recent LOFAR data (see below). As discussed in \citet[][and references therein]{2017ApJ...851....7B}, the low-frequency power-law part is due to supernova ejecta-CSM interaction, and the emission at higher frequencies is presumably from a central object. 
We have included 5 GHz VLBI \citep{2017ApJ...851....7B} and 3 GHz VLASS data \citep{2021ApJ...923L..24S} from 31.6 and 36.1 years, respectively (cf. Fig. \ref{fig:86jflux}), and note that not much evolution is seen in the spectrum of the central component. Within errors, this matches earlier conclusions of \citep{2017ApJ...851....7B} for these frequencies. 

For the low-frequency shell part of the supernova \cite{2017ApJ...851....7B} find a power-law spectral index of $\alpha=0.63\pm0.03$ at 29.1 years, and we obtain $\alpha = 0.58\pm0.03$ at $\sim 30$ years. However, as reported in \cite{Mulcahy18}, the flux for the supernova is estimated by measuring the flux at the position of the source and then subtracting the background emission. This is because the host galaxy is very bright at low radio frequencies in instruments such as LOFAR and GMRT at these resolutions (see Fig. \ref{fig:ngc891_6"}). Since the flux is not directly measured, there may be a level of uncertainty in accurately estimating the background. Further details are not provided for this, so we do not investigate this further. The other flux plotted in Fig. \ref{fig:86jflux} is a LoTSS image at the same resolution of \SI{6}{\arcsec}. However, this measurement is a direct estimate at the position of the supernova and is clearly too high to be of any value. We can compare this with the flux estimated using our ILT image. As seen in Fig. \ref{fig:ngc891_0.5"}, the ILT observation filters out the diffuse flux, resulting in a clear detection of compact sources such as SN 1986J. Although this was taken at a later time of 35.6 years, to compare, we estimate a spectral index of $0.66\pm0.03$ for the shell if combined with the VLA data. It is puzzling that the flux of the supernova should increase with time at lower frequencies, but considering the uncertainty of the LOFAR flux at 30 years, we put less weight on this and trust the ILT flux at 35.6 years more. \cite{2017ApJ...851....7B} find that the low-frequency part decays rapidly, roughly as $\propto t^{-3.92\pm0.07}$ between $15-30$ years. If we extrapolate this to 35.6 years, the shell flux would decrease by a factor of $\sim 2.2$, and the value of $\alpha$ for the shell would be close $\sim 0.98$, which is considerably greater than $\alpha=0.63\pm0.03$. However, if one studies the multiwavelength fit of \cite{2017ApJ...851....7B} for low frequencies, it undershoots by $\sim 30\%$ at 0.3 GHz at 24.4 years, so $\alpha$ should indeed be greater than 0.63 at the lowest frequencies. Further investigation is required by using, for example, GMRT and LOFAR data in conjunction, to estimate the flux and spectral index at low frequencies. 

The very fast decay of the flux of the shell component between $15-30$ years ($\propto t^{-3.92\pm0.07}$) contrasts that of the optically thin emission between $4-6$ years at 5 GHz, which decayed as $\propto t^{-1.19\pm0.03}$ \citep{1990ApJ...364..611W}. The increased decay rate of the shell emission could be a sign that the reverse shock is now in a flatter part of the density profile of the ejecta, as we have argued for to be the case for SN 1979C, that the circumstellar density profile is steeper than $r^{-2}$, or both, as discussed by \cite{2012MNRAS.419.1515D}. This could also explain the vanishing blueshifted part of the broad [\ion{O}{1}] $\lambda6300$, [\ion{O}{2}] $\lambda\lambda7319,7330$, and [\ion{O}{3}] $\lambda5007$ lines between 1991 and 2007 \citep{2008Mili}, which were likely fed by the rapidly decaying X-ray emission \citep[cf.][]{2005Houck} from the reverse shock. Unless the decay of the radio emission from the shell accelerates further, it will stay above 1 mJy at 0.146 GHz roughly until 2040. 

For the central component \cite{2017ApJ...851....7B} obtain good fits to radio emission assuming that the emission suffers from both internal and external thermal absorption \citep[see also][]{2020ApJ...902...55C}. This idea was originally discussed in \cite{1990ApJ...364..611W}. A physical picture for the central component must also explain the narrow optical lines seen already early in the evolution. \cite{1994MNRAS.268..173C} argued that shocked circumstellar clumps are responsible for these lines. \cite{2017ApJ...851....7B} list several likely scenarios that could explain the observations: a common-envelope evolution of the progenitor \citep[originally put forward by][]{2012ApJ...752L...2C}, a pulsar-wind nebula, or accretion onto a newly formed black hole. \cite{2020ApJ...902...55C} noted similarities between the evolution of the radio spectra of SNe 1986J and 2001em, and the authors favor the common-envelope scenario. Shocks driven into the remains of a  hydrogen-rich common-envelope could explain the strong narrow H$\alpha$ emission until 1991, but the rapid decay of this emission until 2007 \citep{2008Mili} would indicate that the shocks should have run through this gas by then. In 2007, the optical spectrum was dominated by broader triangular-shaped [\ion{O}{2}] $\lambda\lambda7319,7330$ and [\ion{O}{3}] $\lambda5007$ lines. They could be from shock-heated ejecta caused by a reverse shock. \cite{2008Mili} note that shocks formed by a pulsar-wind nebula as in the LMC supernova remnant 0540$-$69.3 can also form strong optical forbidden lines. Oxygen-rich filaments deeply embedded in this pulsar-wind nebula are clearly seen in integral-field-unit observations \citep{2013MNRAS.432.2854S,2021ApJ...922..265L}. As modeled by \cite{2023MNRAS.525.2750T}, a pulsar-wind nebula may also contribute to radio emission, but a potential problem for these models is that radio emission increases at least for the first $\sim 100$ years, whereas the opposite is observed for SN 1986J. More observations are needed to discriminate between models for radio, X-ray and optical emission.


 


\section{Conclusions} \label{sec:conclusion}
We presented a LoTSS image of the nearby galaxy M100 and an ILT image of NGC 891 for the first time. For the Type Ia SN 2006X in M100 we find a $3\sigma$ upper limit of $0.7$ mJy at 150 MHz, 12.96 years after the explosion. We have also assembled observations from the other radio facilities e-MERLIN, GMRT and VLA up to the age 14.78 years and derive $3\sigma$ upper limits. We perform a modeling using approximations to the explosion model CS15DD2 by \citet{1999ApJS..125..439I} and compare the predicted radio emission with the observed data. The most sensitive data point is from e-MERLIN at 1.5 GHz at 14.57 years. For the microphysics parameters $\epsilon_{\rm rel}$ and $\epsilon_{\rm B}$ both equal to 0.01, we derive an upper limit on the density of the medium around the supernova to be $\lsim 10\cm3$. Alternatively, for a presumed density of $1\cm3$, this places limits on $\epsilon_{\rm rel}$ and $\epsilon_{\rm B}$; if $\epsilon_{\rm rel} = 0.1$, then $\epsilon_{\rm B} \lesssim 0.021$, or if $\epsilon_{\rm B} = 0.1$, then $\epsilon_{\rm rel} \lesssim 0.019$.

From the LoTSS image of M100, we report a clear detection of the well-studied SN 1979C with a flux density of $4.6\pm0.36$ mJy around 40 years after explosion. We add more data from e-MERLIN, GMRT and VLA up to age 41.62 years. We built a hydrodynamic picture of the supernova using VLBI resolved measurements of the interaction between the supernova and its circumstellar medium together with information from models of the bolometric lightcurve during the first months and then model the radio emission. Our models for the period $22-42$ years point to a zero-age mass of $\lsim 15 \Msun$, probably close to $13 \Msun$. The density profile of the ejecta encountered by the reverse shock, $\rho \propto V_{\rm ej}^n$, during this epoch, when the observed emission around $1.5-1.6$ GHz roughly falls as $F_{\nu} \propto t^{-2.1}$, appears to be described by $n\approx8$. In addition, there is no evidence of the previously suggested continued flattening of the radio spectrum of the supernova after around 2005 \citep{2008ApJ...682.1065B}. Instead, we find progressively steeper spectral indices at later epochs, which contradicts a scenario where emission from a compact object has started to emerge. Extrapolation of our $n=8$ model to earlier epochs revealed that the ejecta density profile close to the reverse shock was significantly steeper at $t \lesssim 20$ years. This is in harmony with a possible increase in radio-structure retardation after $\sim 17$ years. In our models, we use $\dot M_w = 5\times10^{-5}~(v_w/10~\kms)$ $\Msun~{\rm year}^{-1}$. Several solar masses of wind material are likely to remain unshocked by the forward shock. Continued monitoring is encouraged both in radio and X-rays, as well as in the optical to follow the evolution of line strengths and line widths. In particular, it is interesting to monitor any frequency shift of the apparent spectral break at $\sim 1.5$ GHz at 40 years, where the spectrum steepens from approximately $F_{\nu} \propto v^{-0.6}$ to roughly $F_{\nu} \propto v^{-0.94}$. It could be due to synchrotron cooling similar to that found for SN 1993J at $\approx 9$ years \citep{2004ApJ...604L..97C} and could indicate a small $u_{\rm rel}/u_{\rm B}$ ratio of order $10^{-4}$.
Our model with $n=8$ fares well with very recent late X-ray data indicating two components, one decaying with time and with an electron temperature of $0.7-1.1$ keV, and another harder one, which may be thermal or non-thermal and could be near-constant. We identify these as due to X-ray emission from the reverse and forward shocks created by circumstellar interaction.

The ILT image of NGC 891 is capable of filtering out diffuse emission to provide a more accurate estimation of the flux density of $6.77\pm0.2$ mJy SN 1986J at 35.96 years after the explosion. Although the high-frequency VLA data we have are at an age of 29 years, we use them to estimate a spectral index of $0.66\pm0.03$, but taking into account the rapid fall of the low-frequency part of the spectrum, coming from the shell part of the supernova, makes it more likely that the spectral index is closer to $\sim 0.98$. Further investigation is needed to disentangle the differences in flux from LOFAR and the ILT, especially what appears to be a puzzling increase in flux at lower frequencies (which we think may be instrumental). Data from VLASS confirm near constant flux densities for the spectral part, which is because of the central object. This is consistent with previous studies for SN 1986J. We argue that the evolution of narrow optical lines is consistent with shocks related to the central component. The rapidly vanishing emission of H$\alpha$ until 2007 indicates that the hydrogen-rich central component, possibly the remains of a common envelope, had been overtaken by shocks and that the then dominant somewhat broader forbidden oxygen lines are from central shocked supernova ejecta. The oxygen-rich ejecta could be embedded in a pulsar-wind nebula, and this nebula may contribute to high-frequency radio emission. A potential problem is that a model for this predicts an increasing $3-5$ GHz emission \citep{2023MNRAS.525.2750T}, whereas this emission is nearly constant or possibly decreasing. Continued radio, optical, infrared and X-ray observations are encouraged, but we estimate that after 2040, radio emission even at 0.146 GHz from the shell region may fall below 1 mJy.    

We emphasize the importance of observing SNe at late epochs at radio frequencies below a few hundred MHz. As exemplified here by SNe 1979C and 1986J, SNe can remain bright for many decades at such frequencies because of circumstellar interaction. In particular, SN 1979C with its possibly massive still unshocked circumstellar medium is a source that will be persistently bright at low frequencies. At higher frequencies, they are fainter at late times, unless there is significant emission from a central component as in SNe 1986J \citep{2004Sci} and 2001em \citep{2020ApJ...902...55C}. 

\section*{Acknowledgements}
The authors thank Norbert Bartel for discussions and the anonymous referee for important comments. 
This paper is based on data obtained with the International LOFAR Telescope (ILT) under project code LC10\_010 and LoTSS. LoTSS Data Release 2 data are described by \cite{2022A&A...659A...1S}.
LOFAR data products were provided by the LOFAR Surveys Key Science project (LSKSP; \url{https://lofar-surveys.org/}) and were derived from observations with the International LOFAR Telescope (ILT). LOFAR \citep{2013A&A...556A...2V} is the Low Frequency Array designed and constructed by ASTRON. It has observing, data processing, and data storage facilities in several countries, which are owned by various parties (each with their own funding sources) and which are collectively operated by the ILT foundation under a joint scientific policy. The efforts of the LSKSP have benefited from funding from the European Research Council, NOVA, NWO, CNRS-INSU, the SURF Co-operative, the UK Science and Technology Funding Council and the Jülich Supercomputing Centre. This research used the NASA/IPAC Extragalactic Database (NED), which is funded by the National Aeronautics and Space Administration and operated by the California Institute of Technology.
This work used GMRT data. The authors thank the GMRT staff that made these observations possible. GMRT is run by the National Centre for Radio Astrophysics of the Tata Institute of Fundamental Research.
This work also used e-MERLIN data. e-MERLIN is a National Facility operated by the University of Manchester at Jodrell Bank Observatory on behalf of STFC.
MPT and JM acknowledge financial support from the Severo Ochoa grant CEX2021-001131-S and from the Spanish grant PID2023-147883NB-C21, funded by MCIU/AEI/ 10.13039/501100011033, as well as support through ERDF/EU.

\facilities{LOFAR, VLA, e-MERLIN, GMRT}

\software{astropy \citep{2022ApJ...935..167A}, 
          \textsc{APLpy} (\citealt{aplpy2012,aplpy2019}),
          \textsc{NumPy} \citep{harris2020array},
          \textsc{SciPy} \citep{2020SciPy-NMeth},
          \textsc{Matplotlib} \citep{Hunter:2007},
          \textsc{CASA} \citep{2022PASP..134k4501C},
          \textsc{LINC} \citep{2019A&A...622A...5D},
          \textsc{DDFacet} \citep{2023ascl.soft05008T},
          \textsc{e-MERLIN CASA pipeline}
          \citep{2021ascl.soft09006M},  
          LOFAR-VLBI pipeline \citep{2022A&A...658A...1M}
          }



\bibliography{sample631}{}
\bibliographystyle{aasjournal}



\end{document}